\def \VersionLong {}
\ifdefined \VersionLong
	\newcommand{\LongVersion}[1]{\ifdefined\VersionWithComments{\color{red!40!black}#1}\else#1\fi}
	\newcommand{\ShortVersion}[1]{\ifdefined\VersionWithComments{\color{black!40}#1}\fi}
\else
	\newcommand{\LongVersion}[1]{\ifdefined\VersionWithComments{\color{black!40}#1}\fi}
	\newcommand{\ShortVersion}[1]{\ifdefined\VersionWithComments{\color{red!40!black}#1}\else#1\fi}
\fi

\documentclass[a4paper,10pt]{llncs}
\makeatletter
\AtBeginDocument{%
  \@ifpackageloaded{hyperref}
  {\def\@doi#1{\href{https://doi.org/#1}
      {\ttfamily https://doi.org/#1}\egroup}}
  {\def\@doi#1{\ttfamily https://doi.org/#1\egroup}}
  \def\doi{\bgroup\catcode`\_=12\relax\@doi}}
\makeatother
\usepackage[utf8]{inputenc}
\usepackage[english]{babel}

\usepackage{setspace}

\usepackage[ruled,lined,linesnumbered,noend]{algorithm2e}
\SetKwInOut{Input}{input}
\SetKwInOut{Output}{output}
\SetKw{KwPush}{push}
\SetKw{KwPop}{pop}
\SetKw{KwFrom}{from}
\SetKw{KwCompute}{compute}

\usepackage{subcaption}
\usepackage{paralist} % inline lists
\usepackage{xspace}

\newenvironment{ienumeration}
	{\ifdefined\VersionLong\begin{enumerate}\else\begin{inparaenum}[\itshape i\upshape)]\fi}
	{\ifdefined\VersionLong\end{enumerate}\else\end{inparaenum}\fi}

\usepackage{amsmath} % for \text et al., for ``cases'' environment…
\usepackage{amssymb} % for \mathbb et al.
\usepackage{multirow}

\usepackage[misc,geometry]{ifsym} % For \Letter

\usepackage{eurosym} % FOR EURO SYMBOL

\usepackage{listings}
\usepackage{color}

	\definecolor{mygreen}{rgb}{0,0.6,0}
	\definecolor{mygray}{rgb}{0.5,0.5,0.5}
	\definecolor{mymauve}{rgb}{0.58,0,0.82}
	\definecolor{weborange}{RGB}{255,165,0}
\lstdefinestyle{log}{
	backgroundcolor=\color{white},   % choose the background color; you must add \usepackage{color} or \usepackage{xcolor}; should come as last argument
	basicstyle=\scriptsize,        % the size of the fonts that are used for the code
	breakatwhitespace=false,         % sets if automatic breaks should only happen at whitespace
	breaklines=true,                 % sets automatic line breaking
	captionpos=b,                    % sets the caption-position to bottom
	commentstyle=\color{mygreen},    % comment style
	deletekeywords={...},            % if you want to delete keywords from the given language
	escapeinside={\%*}{*)},          % if you want to add LaTeX within your code
	extendedchars=true,              % lets you use non-ASCII characters; for 8-bits encodings only, does not work with UTF-8
	frame=single,	                   % adds a frame around the code
	keepspaces=true,                 % keeps spaces in text, useful for keeping indentation of code (possibly needs columns=flexible)
	keywordstyle=\color{red!70!black}\bfseries,       % keyword style
	morekeywords={@, open, close, update},            % if you want to add more keywords to the set
	numbers=left,                    % where to put the line-numbers; possible values are (none, left, right)
	numbersep=5pt,                   % how far the line-numbers are from the code
	numberstyle=\tiny\color{mygray}, % the style that is used for the line-numbers
	rulecolor=\color{black},         % if not set, the frame-color may be changed on line-breaks within not-black text (e.g. comments (green here))
	showspaces=false,                % show spaces everywhere adding particular underscores; it overrides 'showstringspaces'
	showstringspaces=false,          % underline spaces within strings only
	showtabs=false,                  % show tabs within strings adding particular underscores
	stepnumber=1,                    % the step between two line-numbers. If it's 1, each line will be numbered
	stringstyle=\color{mymauve},     % string literal style
	tabsize=2,	                   % sets default tabsize to 2 spaces
	classoffset=1, % starting new class
	otherkeywords={@},
	morekeywords={@},
	keywordstyle=\color{weborange},
	classoffset=0,
}

\ifdefined\VersionWithComments
	\usepackage{draftwatermark}
	\SetWatermarkText{draft}
	\SetWatermarkScale{15}
	\SetWatermarkColor[gray]{0.9}
\fi
\usepackage[svgnames,table]{xcolor}
\definecolor{darkblue}{rgb}{0.0,0.0,0.6}
\definecolor{darkgreen}{rgb}{0, 0.5, 0}
\definecolor{darkpurple}{rgb}{0.7, 0, 0.7}
\definecolor{darkblue}{rgb}{0, 0, 0.7}

\usepackage[
		colorlinks=true,
	\ifdefined \VersionLong
		pagebackref=true,
	\fi
		citecolor=darkgreen,
		linkcolor=darkblue,
		urlcolor=darkpurple,
	]{hyperref}
	
\usepackage[capitalise,english,nameinlink]{cleveref} % load after algorithm2e and hyperref
\crefname{line}{\text{line}}{\text{lines}} % to remove the capital
\crefname{assumption}{\text{Assumption}}{\text{Assumptions}} % added Assumption

\usepackage{wrapfig}
\usepackage{tikz}
\usetikzlibrary{arrows,automata,positioning}
\tikzstyle{every node}=[initial text=]
	\tikzstyle{location}=[circle, minimum size=12pt, draw=black, fill=blue!10, inner sep=1pt] % rectangle, rounded corners
\tikzstyle{final}=[double]
\tikzstyle{accepting}=[final]
\tikzstyle{PTPMOPT}=[,dashed,color=red,semithick]

\newcommand{\styleact}[1]{\ensuremath{\textcolor{coloract!80!black}{\mathrm{#1}}}}
\newcommand{\styleclock}[1]{\ensuremath{\textcolor{colorclock!80!black}{#1}}}

\newcommand{\styleparam}[1]{\ensuremath{\textcolor{colorparam!80!black}{\mathsf{#1}}}}
\newcommand{\stylelvar}[1]{\ensuremath{\textcolor{colorlvar!80!black}{#1}}}
\newcommand{\stylevar}[1]{\ensuremath{\textcolor{colorvar!80!black}{#1}}}
\newcommand{\styledparam}[1]{\ensuremath{\textcolor{colordparam!80!black}{\mathsf{#1}}}}

	\definecolor{coloract}{rgb}{0.50, 0.70, 0.30}
	\definecolor{colorclock}{rgb}{0.4, 0.4, 1}
	\definecolor{colorconst}{rgb}{0.50, 0.20, 0.00}
	\definecolor{colordisc}{rgb}{1, 0, 1}
	\definecolor{colorloc}{rgb}{0.4, 0.4, 0.65}
	\definecolor{colorparam}{rgb}{1, 0.6, 0.0}
	\definecolor{colorvar}{rgb}{0.6, 0.7, 1}
	\definecolor{colorlvar}{rgb}{0.4, 0.4, .5}
	\definecolor{colordparam}{rgb}{.9, 0.8, 0.0}
\usepackage{wasysym}

\newcommand{\smileyunhappy}{\bf\textcolor{red}\frownie}

\newif\iftikzgnuplot
\tikzgnuplottrue 
\usepackage{gnuplot-lua-tikz}
\usepackage{pgfplotstable}
\pgfplotsset{compat=1.12}
\usepackage{booktabs}

\newcommand{\init}{_0}

\newcommand{\A}{\ensuremath{\mathcal{A}}}
\newcommand{\Actions}{\Sigma}
\newcommand{\ActDomain}{\ensuremath{\mathit{Dom}}}
\newcommand{\action}{\ensuremath{a}}
\newcommand{\ArExp}{\mathcal{LA}} % linear arithmetic expressions

\newcommand{\Clock}{\mathbb{C}} % set of clocks
\newcommand{\ClockCard}{H} % cardinality of clocks
\renewcommand{\clock}{c} % clock % NOTE: due to package wasysym
\newcommand{\clockval}{\nu} % clock valuation
\newcommand{\ClocksZero}{\vec{0}}
\newcommand{\compOp}{\bowtie}
\newcommand{\compString}{\approx}
\newcommand{\CVal}{({\Rgeqzero})^{\Clock}}

\newcommand{\Domain}{\mathbb{D}}
\newcommand{\DomainUpdates}{\mathcal{DU}}
\newcommand{\DomainGuards}{\mathcal{DE}}
\newcommand{\DomainVar}{\ensuremath{\mathit{Var}}}
\newcommand{\dpval}{\ensuremath{\zeta}}
\newcommand{\edge}{e}
\newcommand{\Edges}{E}
\newcommand{\longueflecheRel}[1]{\stackrel{#1}{\mapsto}}

\newcommand{\flecheRel}{{\rightarrow}}

\newcommand{\freeze}{*}
\newcommand{\grandn}{{\mathbb N}}
\newcommand{\grandq}{{\mathbb Q}}
\newcommand{\grandqplus}{\grandq_{+}} % \geq 0
\newcommand{\grands}{{\mathbb S}}
\newcommand{\grandz}{{\mathbb Z}}
\newcommand{\tguard}{tg}
\newcommand{\dguard}{dg}

\newcommand{\Lg}{\mathcal{L}}
\newcommand{\loc}{\ell} % location
\newcommand{\locinit}{\loc\init}
\newcommand{\Loc}{L} % set of locations
\newcommand{\LocFinal}{F}
\newcommand{\TParam}{\mathbb{TP}} % set of timing parameters (P)
\newcommand{\tparam}{\ensuremath{\mathsf{tp}}} % timing parameter (p)
\newcommand{\TParamCard}{J} % number of parameters
\newcommand{\PVal}{({\grandqplus})^{\TParam}}
\newcommand{\tpval}{\ensuremath{\gamma}} % parameter valuation
\newcommand{\PDU}{\ensuremath{\mathsf{PDU}}}

\newcommand{\R}{{\mathbb{R}}}
\newcommand{\Rgeqzero}{\R_{\geq 0}}

\newcommand{\sinit}{s\init} % initial set of states
\newcommand{\States}{S} % for LTS
\newcommand{\LVariables}{\mathbb{LV}}
\newcommand{\lvar}{\mathit{lv}}
\newcommand{\lvarval}{\eta}
\newcommand{\LVariablesCard}{O}

\newcommand{\Variables}{\mathbb{V}}
\newcommand{\varval}{\mu}
\newcommand{\varvalo}{\varval\init}
\newcommand{\dvar}{v}
\newcommand{\VariablesCard}{M}
\newcommand{\VParam}{\mathbb{VP}}
\newcommand{\vparam}{\mathsf{vp}}
\newcommand{\VParamCard}{N} % number of parameters
\newcommand{\word}{\textcolor{colorok}{w}}
\newcommand{\words}{\textcolor{colorok}{W}}
\newcommand{\DPValuate}[4]
	{\ensuremath{\reset{#3}{\valuate{\valuate{#1}{#2}}{#4}}}}  % 1: update function 2: data parameter valuation 3: data (variable) valuation 4: local variable valuation

\newcommand{\partfun}{\nrightarrow}

\newcommand{\resets}{R}
\newcommand{\project}[2]{\ensuremath{#1{\downarrow_{#2}}}}
\newcommand{\reset}[2]{\ensuremath{[#1]_{#2}}}
\newcommand{\valuate}[2]{\ensuremath{#2(#1)}}
\newcommand{\dtpval}[2]{#1|#2} % (w,v)

\newcommand{\Rp}{{\mathbb{R}_{>0}}}
\newcommand{\Rnn}{{\Rgeqzero}}
\newcommand{\disjointUnion}{\sqcup}

\newcommand{\finitePowerset}[1]{\mathcal{P}_{\mathrm{fin}}(#1)}

\usepackage{pifont}% http://ctan.org/pkg/pifont

\newcommand{\CurrUConf}{\mathit{CurrUConf}}
\newcommand{\CurrConf}{\mathit{CurrConf}}

\newcommand{\NextConf}{\mathit{NextConf}}

\newcommand{\uAction}{\varepsilon}
\newcommand{\ActionsWithU}{\Actions_{\uAction}}

\newcommand{\ValidityDomain}[2]{D(#1,#2)}

\newcommand{\clocktpvals}{V_{t}}
\newcommand{\clocktpvalsDwell}{V_{t+}}
\newcommand{\vardpvals}{V_{d}}
\newcommand{\Result}{\mathit{Result}}
\newcommand{\Resulti}[1]{\mathit{Result}_{#1}}
\newcommand{\UnobservableConf}[1]{\mathit{Conf}^{u}_{#1}}
\newcommand{\ObservableConf}[1]{\mathit{Conf}^{o}_{#1}}

\newcommand{\datatype}{data type}

\newcommand{\defProblem}[3]
{%	
\noindent\fcolorbox{black}{blue!15}{ % FOR ANONYMOUS VERSION
	\begin{minipage}{.95\columnwidth}
		\textbf{#1 problem:}\\
		\textsc{Input}: #2\\
		\textsc{Problem}: #3
	\end{minipage}
}
}

\definecolor{vertfonce}{rgb}{0.0, 0.5, 0.0}
\definecolor{rougefonce}{rgb}{1, 0.0, 0.0}
\newcommand{\compyes}{$\textcolor{vertfonce}{\mathbf{\surd}}$}
\newcommand{\compno}{$\textcolor{rougefonce}{\mathbf{\times}}$}

\newcommand{\cellYes}{\cellcolor{green!20}\textbf{\compyes}}
\newcommand{\cellYesNo}{\cellcolor{orange!20}\textbf{\compyes/\compno}}
\newcommand{\cellNo}{\cellcolor{red!20}\textbf{\compno}}
\newcommand{\cellNA}{\cellcolor{black!20}N/A}
\newcommand{\cellUnclear}{\cellcolor{yellow!20}\textbf{?}}

	\definecolor{cellcolor}{rgb}{.8, .8, 1}
\newcommand{\rowHeader}{\rowcolor{cellcolor}\bfseries}

\newtheorem{assumption}{Assumption}
\usepackage{marginnote}
\ifdefined \VersionWithComments
	\newcommand{\marginX}{\marginnote{\huge{\quad\quad\textbf{!}\quad\quad}}}
	\newcommand{\ea}[1]{\mbox{}{\color{blue}\marginX{}\textbf{[\'Etienne}: #1]}}
	\newcommand{\ih}[1]{\mbox{}{\color{purple}\marginX{}\textbf{[Ichiro}: #1]}}
	\newcommand{\mw}[1]{\mbox{}{\color{orange}\marginX{}\textbf{[Masaki}: #1]}}
	\newcommand{\instructions}[1]{{\color{red}\marginX{}\textbf{[Instructions: ``#1'']}}}
	\newcommand{\reviewer}[2]{\mbox{}{\color{red}\marginX{}\textbf{[Reviewer #1}: ``#2'']}}
	\newcommand{\todo}[1]{\mbox{}{\color{red}{\marginX{}\textbf{TODO}\ifx#1\\\else:\ \fi #1}}} % here, ``\\'' stands for ``empty''
\else
	\newcommand{\instructions}[1]{}
	\newcommand{\ea}[1]{}
	\newcommand{\ih}[1]{}
	\newcommand{\mw}[1]{}
	\newcommand{\reviewer}[2]{}
	\newcommand{\todo}[1]{}
\fi

\newcommand{\DejaVu}{\textsc{DejaVu}}
\newcommand{\imitator}{\textsf{IMITATOR}}
\newcommand{\MonPoly}{\textsc{MonPoly}}

\newcommand{\masakiTool}{\textsc{SyMon}}

\ifdefined \VersionWithComments
 	\definecolor{colorok}{RGB}{80,80,150}
\else
	\definecolor{colorok}{RGB}{0,0,0}
\fi

\newcommand{\eg}{\textcolor{colorok}{e.\,g.,}\xspace}
\newcommand{\ie}{\textcolor{colorok}{i.\,e.,}\xspace}
\newcommand{\st}{\textcolor{colorok}{s.t.}\xspace}

\newcommand{\wrt}{\textcolor{colorok}{w.r.t.}\xspace}

\title{Symbolic Monitoring against Specifications Parametric in Time and Data\ifdefined \VersionWithComments
	\textcolor{red}{{ [Version: \today{}]}}
\fi
 \thanks{%
 	\LongVersion{%
 		This is the author version of the manuscript of the same name published in the proceedings of the 31st International Conference on Computer-Aided Verification (CAV 2019).
 	        The final version is available at \url{www.springer.com}.
 	}%
 	This work is partially supported 
 	by
 	JST ERATO HASUO Metamathematics for Systems Design Project (No.\ JPMJER1603),
         by JSPS Grants-in-Aid No.\ 15KT0012 \& 18J22498
         and 
        by the ANR national research program PACS (ANR-14-CE28-0002).
 }
}
 \author{Masaki Waga\inst{1,2,3}\orcidID{0000-0001-9360-7490}\Letter{} \and \'Etienne Andr\'e\inst{4,5,1}\orcidID{0000-0001-8473-9555} \and Ichiro Hasuo\inst{1,2}\orcidID{0000-0002-8300-4650}}
 \date{\today{}}
 \institute{%
 National Institute of Informatics, Tokyo, Japan
 \and
 SOKENDAI (The Graduate University for Advanced Studies), Tokyo, Japan
 \and
 JSPS Research Fellow
 \and
 Université Paris 13, LIPN, CNRS, UMR 7030, F-93430, Villetaneuse, France
 \and
 JFLI, CNRS, Tokyo, Japan
 }

\usepackage{graphicx}	% required for `\scalebox' (yatex added)
\usepackage{version}	% required for `\comment' (yatex added)
\begin{document}

% For all page numbers, except p.1
\LongVersion{
\pagestyle{plain}
}

\maketitle

% HACK: because of LNCS style!
\setcounter{footnote}{0}

% For page numbers on p.1
\LongVersion{
\thispagestyle{plain}
}

\ifdefined \VersionWithComments
	\textcolor{red}{\textbf{This is the version with comments. To disable comments, comment out line~3 in the \LaTeX{} source.}}
\fi

\begin{abstract}
	Monitoring consists in deciding whether a log meets a given specification.
	In this work, we propose an automata-based formalism to monitor logs in the form of actions associated with time stamps and arbitrarily data values over infinite domains.
	Our formalism uses both timing parameters and data parameters, and is able to output answers symbolic in these parameters and in the log segments where the property is satisfied or violated.%\mw{I do not understand this part well. Symbolic monitoring outputs the validity domain (parameter valuations) but it is not something like pattern matching, though PTPM can be encoded using timing parameters.}\ea{True. What about ``Our formalism uses both timing parameters and data parameters, and is able to output answers symbolic in these parameters for which the property is satisfied or violated. We also show that our approach is sufficiently expressive to encode parametric timed pattern matching.''}
	We implemented our approach in an ad-hoc prototype \masakiTool{}, and experiments show that its high expressive power still allows for efficient online monitoring.
	
	\LongVersion{}
\end{abstract}

\instructions{Regular Papers should not exceed 16 pages in LNCS format, not counting references and appendices.}

% \todo{cite ``First order temporal logic monitoring with BDDs''! \cite{HPU17} (and add for our data?!)}

% \ea{hello}
% \ih{hello}
% \mw{hello}

%%%%%%%%%%%%%%%%%%%%%%%%%%%%%%%%%%%%%%%%%%%%%%%%%%%%%%%%%%%%
%%%%%%%%%%%%%%%%%%%%%%%%%%%%%%%%%%%%%%%%%%%%%%%%%%%%%%%%%%%%
\section{Introduction}\label{section:introduction}
%%%%%%%%%%%%%%%%%%%%%%%%%%%%%%%%%%%%%%%%%%%%%%%%%%%%%%%%%%%%
%%%%%%%%%%%%%%%%%%%%%%%%%%%%%%%%%%%%%%%%%%%%%%%%%%%%%%%%%%%%

Monitoring consists in checking whether a sequence of data (a log or a signal) satisfies or violates a specification expressed using some formalism.
Offline monitoring consists in performing this analysis after the system execution, as the technique has access to the entire log in order to decide whether the specification is violated.
In contrast, online monitoring can make a decision earlier, ideally as soon as a witness of the violation of the specification is encountered.

Using existing formalisms (\eg{} the metric first order temporal logic~\cite{BKMZ15}), one can check whether a given bank customer withdraws more than 1,000\,\euro{} every week.
With formalisms extended with data, one may even \emph{identify} such customers.
Or, using an extension of the signal temporal logic (STL)~\cite{BDSV14}, one can ask: ``is that true that the value of variable~$x$ is always copied to~$y$ exactly 4~time units later?''
However, questions relating time and data using parameters become much harder (or even impossible) to express using existing formalisms:
``what are the users and time frames during which a user withdraws more than half of the total bank withdrawals within seven days?''
And even, can we \emph{synthesize} the durations (not necessarily 7~days) for which this specification holds?
Or
``what is the set of variables for which there exists a duration within which their value is always copied to another variable?''
In addition, detecting periodic behaviors without knowing the period can be hard to achieve using existing formalisms.

In this work, we address the challenging problem to monitor logs enriched with both timing information and (infinite domain) data.
In addition, we significantly push the existing limits of expressiveness so as to allow for a further level of abstraction using \emph{parameters}:
	our specification can be both parametric in the \emph{time} and in the \emph{data}.
The answer to this symbolic monitoring is richer than a pure Boolean answer, as it \emph{synthesizes} the values of both time and data parameters for which the specification holds.
This allows us notably to detect periodic behaviors without knowing the period while being symbolic in terms of data.
For example, we can \emph{synthesize} \emph{variable names} (data) and \emph{delays} for which variables will have their value copied to another data within the aforementioned delay.
In addition, we show that we can detect the log \emph{segments} (start and end date) for which a specification holds.

\begin{figure*}[t]
	%----------------------------------------------------------
	\begin{subfigure}[b]{.42\textwidth}
	\centering
		\footnotesize
	\begin{lstlisting}[style=log]
@0 update(a,0)  @4 update(c,2)
@1 update(c,1)  @5 update(a,2)
@2 update(a,0)  @6 update(b,2)
@3 update(b,1)  @7 update(c,3)
@4 update(b,0)  @9 update(b,3)
\end{lstlisting}
		
	\caption{Log}
	\label{figure:examples:copy:log}
	\end{subfigure}
	%----------------------------------------------------------
	%
	%----------------------------------------------------------
	\begin{subfigure}[b]{.52\textwidth}
	\centering
		\footnotesize
		
		\scalebox{1}{
		\begin{tikzpicture}[shorten >=1pt, scale=.5, yscale=.7, auto]
% 		\node[location] (lgood) at (3, -1) {$=$};

		% Axes
		\draw[->] (0, 0) --++ (0, 3.5);
		\draw[->] (0, 0) --++ (10.5, 0) node[anchor=north]{$t$};
		\foreach \x in {0, 1, ..., 9} % X
			\draw (\x, 0) -- (\x, -.2)node [below] {\scriptsize{$\x$}};
		
% 		%% edges
		\draw[thick, blue] (0, 0) -- (5, 0) -- (5, 2) -- (10, 2) node[anchor=north]{$\mathtt{a}$};
		\draw[thick, dashed, green!50!black] (0, 0) -- (1, 0) -- (1, 1) -- (4, 1) -- (4, 2) -- (7, 2) -- (7, 3) -- (10, 3) node[anchor=south]{$\mathtt{c}$};
		\draw[thick, densely dotted, red] (0, 0) -- (3, 0) -- (3, 1) -- (4, 1) -- (4, 0) -- (6, 0) -- (6, 2) -- (9, 2) -- (9, 3) -- (10, 3) node[anchor=north]{$\mathtt{b}$};

		\end{tikzpicture}
		}		
	\caption{Graphical representation}
	\label{figure:examples:copy:signal}
	\end{subfigure}
	%----------------------------------------------------------

	%----------------------------------------------------------
	\begin{subfigure}[b]{\textwidth}
	\centering
		
		\scalebox{.75}{
		\begin{tikzpicture}[shorten >=1pt, scale=2, yscale=1, auto] % node distance=4.5cm and 2cm node distance=3cm and 1cm, 
		%% states
		\node[location,initial] at (0, 0) (l0) {$\loc_0$};
		\node[location] (l1) at (3, 0) {$\neq$};
		\node[location,accepting] (lbad) at (5, -1) {$\loc_{\smileyunhappy}$};
		\node[location] (lgood) at (3, -1) {$=$};

% 		%% edges
		\path[->] 
		(l0) edge [loop above] node[align=center] {$\styleact{update}(\stylelvar{x}, \stylelvar{v})$ \\ $\stylelvar{x} \neq \texttt{b}$} (l0)
		(l0) edge [loop below] node[align=center] {$\styleact{update}(\texttt{b}, \stylelvar{v})$ \\ $\stylevar{valb} := \stylelvar{v}$} (l0)
		
		(l0) edge [] node[below, align=center, xshift=-1.5em] {$\styleact{update}(\stylelvar{x}, \stylelvar{v})$ \\ $\stylelvar{x} = \styledparam{px}$\\ $\stylevar{valb} = \stylelvar{v}$} (lgood)
		
		(l0) edge [] node[align=center] {$\styleact{update}(\stylelvar{x}, \stylelvar{v})$ \\ $\stylelvar{x} = \styledparam{px}$\\ $\stylevar{valb} \neq \stylelvar{v}$ \\ $\styleclock{\clock} := 0$, $\stylevar{valx} := \stylelvar{v}$} (l1)
		
		(l1) edge [] node[left, align=center, yshift=.5em] {$\styleact{update}(\texttt{b}, \stylelvar{v})$ \\ $\stylelvar{v} = \stylevar{valx}$\\ $\styleclock{\clock} < \styleparam{\tparam}$} (lgood)
		
		(l1) edge [] node[below, align=center] {$\styleact{\uAction}$ \\ $\styleclock{\clock} \geq \styleparam{\tparam}$} (lbad)
		
		(l1) edge [loop right] node[right, align=center] {$\styleact{update}(\texttt{b}, \stylelvar{v})$ \\ $\styleclock{\clock} < \styleparam{\tparam}$ \\ $\stylelvar{v} \neq \stylevar{valx}$} (l1)
		
		(l1) edge [loop above] node[above right, xshift=-.5em, yshift=-1em, align=center] {$\styleact{update}(\stylelvar{x}, \stylelvar{v})$ \\ $\styleclock{\clock} < \styleparam{\tparam}$ \\ $\stylelvar{x} \neq \texttt{b}$} (l1)
	;
		\end{tikzpicture}
		}
	
	\caption{Monitoring PTDA}
	\label{figure:example:copy:automaton}
	
	\end{subfigure}
	%----------------------------------------------------------
	
	\caption{Monitoring copy to \texttt{b} within \styleparam{\tparam} time units}
	\label{figure:example:copy}
\end{figure*}
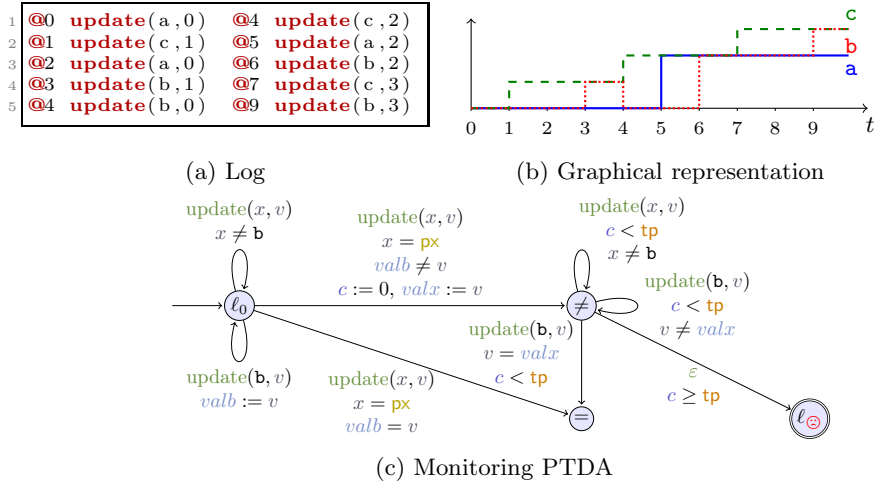

%----------------------------------------------------------
\begin{example}\label{example:copy}
	Consider a system updating three variables \texttt{a}, \texttt{b} and \texttt{c} (\ie{} strings) to values (rationals).
	An example of log is given in \cref{figure:examples:copy:log}.
	Although our work is event-based, we can give a graphical representation similar to that of signals in \cref{figure:examples:copy:signal}.
	Consider the following property: ``for any variable~$\styledparam{px}$, whenever an update of that variable occurs, then within strictly less than $\styleparam{\tparam}$ time units, the value of variable~\texttt{b} must be equal to that update''.\LongVersion{ In our formalism, a simple automaton made of 4 locations (given in \cref{figure:example:copy:automaton}) can monitor this property.}
The \emph{variable parameter}~$\styledparam{px}$ is compared with string values and the \emph{timing parameter}~$\styleparam{\tparam}$ is used in the timing constraints. 
%\todo{Make it clearer what variable/timing parameters are.}
 We are interested in checking %for which segments of the log,
		for which values of\LongVersion{ the variable parameter}~$\styledparam{px}$ and\LongVersion{ the timing parameter} $\styleparam{\tparam}$ this property is violated.
	This can be seen as a synthesis problem in both the variable and timing parameters.
	For example, $\styledparam{px} = \mathtt{c}$ and $\styleparam{\tparam} = 1.5$ %\ea{I use an abuse of notation instead of $\tpval(\styleparam{\tparam}) = 1.5$; is that OK?} 
 is a violation of the specification, as the update of~$\mathtt{c}$ to~2 at time~4 is not propagated to~$\mathtt{b}$ within~1.5 time unit.
 Our algorithm outputs such violation by a constraint \eg{} $\styledparam{px} = c \land \styleparam{\tparam} \leq 2$.
	In contrast, the value of any signal at any time is always such that either \texttt{b} is equal to that signal, or the value of \texttt{b} will be equal to that value within at most 2 time units.
	Thus, the specification holds for any valuation of the variable parameter~$\styledparam{px}$, provided $\styleparam{\tparam} > 2$.
% \reviewer{2}{Not sure I understand these two lines.}
% \todo{The idea is that the value of any signal at any time is always such that either b is equal to that signal, or the value of b will be equal to that value within at most 2 time units. We will add more details. (Note that >= 2 should be > 2 though.)}
%	\todo{for now, \cref{figure:example:copy:automaton} is not used (but it's used in the experiments)}\ea{Perhaps mention ``In our formalism, a simple automaton made of 4 locations (given in \cref{figure:example:copy:automaton}) can monitor this property.''}
%\todo{shown an example of our output like \cref{example:symbolic_monitoring}}
\mw{I updated to clarify the unclear points for the reviewers.}
\end{example}
%----------------------------------------------------------

%----------------------------------------------------------
%----------------------------------------------------------

We propose an automata-based approach to perform monitoring parametric in both time and data.
\LongVersion{%
	We use an extension of both timed automata extended with data, and of parametric timed automata, both extended with parametric data over infinite domains.
}%
We implement our work in a\LongVersion{n ad-hoc} prototype \masakiTool{}\LongVersion{ (relying on polyhedra to encode symbolic parameter relations)} and perform experiments showing that, while our formalism allows for high expressiveness, it is also tractable even for online monitoring.

%----------------------------------------------------------
\begin{table}[tbh]
	\centering

	\caption{Comparison of monitoring expressiveness}%\small
	\LongVersion{\scriptsize}
	\ShortVersion{\scriptsize}
	\begin{tabular}{ l | c | c | c | c | c | c | c | c | c |}
% 		\hline
	\rowHeader{}
		Work
			& \cite{ADMN11} % parametric identification of temporal properties
			& \cite{BDSV14} % STL*
			& \cite{BKMZ15} % MFOTL
			& \cite{BKMZ15FMSD} % MFOTL (+ aggregations)
			& \cite{RCR15} % automata-based with data parameters
			& \cite{HPU17} % symbolic data with BDDs
			& \cite{AHW18} % our work ICECCS 2018
			& \cite{BFM18} % parameter identification for STL
			& This work
		\\
	%	                                  & ADMN11         & BDSV14         &    BKMZ15      &    BKMZ15FMSD  & RCR15          &    HPU17       &     AHW18      &    BFM18       & US
% 		\hline
% 		Timed answer           & \cellYes{}     & \cellNo{}      & \cellNo{}      & \cellNo{}      & \cellNo{}      & \cellNo{}      & \cellYes{}     & \cellUnclear{} & \cellYes{}
% 		\\
		\hline
		Timing parameters                 & \cellYes{}     & \cellNo{}      & \cellUnclear{} & \cellUnclear{} & \cellUnclear{} & \cellNo{}      & \cellYes{}     & \cellNo{}      & \cellYes{}
		\\
		\hline
		Data                              & \cellYes{}     & \cellYes{}     & \cellYes{}     & \cellYes{}     & \cellYes{}     & \cellYes{}     & \cellNo{}      & \cellYes{}     & \cellYes{}
		\\
		\hline
		Parametric data                   & \cellYes{}     & \cellNo{}      & \cellYes{}     & \cellYes{}     & \cellYes{}     & \cellYes{}     & \cellNo{}      & \cellYes{}     & \cellYes{}
		\\
		\hline
		Memory                            & \cellNo{}      & \cellYes{}     & \cellYes{}     & \cellYes{}     & \cellYes{}     & \cellYes{}     & \cellNo{}      & \cellNo{}      & \cellYes{}
		\\
		\hline
		Aggregation                       & \cellNo{}      & \cellNo{}      & \cellNo{}      & \cellYes{}     & \cellYes{}     & \cellNo{}      & \cellNo{}      & \cellNo{}      & \cellYes{}
		\\
		\hline
% 		Symbolic answer        & \cellYes{}     & \cellNo{}      & \cellNo{}      & \cellNo{}      & \cellNo{}      & \cellYes{}     & \cellYes{}     & \cellYes{}     & \cellYes{}
% 	%	                                  & \cellUnclear{} & \cellUnclear{} & \cellUnclear{} & \cellUnclear{} & \cellUnclear{} & \cellUnclear{} & \cellUnclear{} & \cellUnclear{} & \cellYes{}
% 		\\
% 		\hline
		Complete parameter identification & \cellYes{}     & \cellNA{}      & \cellYesNo{}   & \cellYesNo{}   & \cellNA{}      & \cellNA{}      & \cellYes{}     & \cellYes{}     & \cellYes{}
		\\
		\hline
	\end{tabular}

	\label{table:comparisons}
\end{table}
%----------------------------------------------------------

% BEGIN NEW TEXT BY ICHIRO
% a. expressiveness-performance balance
%   i
We believe our framework balances expressiveness and monitoring performance well:
\begin{ienumeration}
	\item Regarding expressiveness, comparison with the existing work is summarized in \cref{table:comparisons} (see \cref{section:related} for further details).
	\item Our monitoring is \emph{complete}, in the sense that it returns a symbolic constraint characterizing \emph{all} the parameter valuations that match a given specification.
	\item We also achieve reasonable monitoring speed, especially given the degree of parametrization in our formalism.
\end{ienumeration}
%   ii
Note that it is not easy to formally claim superiority in expressiveness: proofs would require arguments such as the pumping lemma; and such formal comparison does not seem to be a concern of the existing work.
Moreover, such formal comparison bears little importance for industrial practitioners: expressivity via an elaborate encoding is hardly of practical use.
% b. Coincidence of expressiveness between theory and tool/implementation. 
We also note that, in the existing work, we often observe gaps between the formalism in a theory and the formalism that the resulting tool actually accepts.
This is not the case with the current framework.
% END NEW TEXT BY ICHIRO

%%%%%%%%%%%%%%%%%%%%%%%%%%%%%%%%%%%%%%%%%%%%%%%%%%%%%%%%%%%%
\paragraph{Outline}
%%%%%%%%%%%%%%%%%%%%%%%%%%%%%%%%%%%%%%%%%%%%%%%%%%%%%%%%%%%%
After discussing related works in \cref{section:related},
we introduce the necessary preliminaries in \cref{section:preliminaries}, and our parametric timed data automata in \cref{section:formalism}.
We present our symbolic monitoring approach in \cref{section:monitoring} and conduct experiments in \cref{section:experiments}.
We conclude in \cref{section:conclusion}.

%%%%%%%%%%%%%%%%%%%%%%%%%%%%%%%%%%%%%%%%%%%%%%%%%%%%%%%%%%%%
\section{Related works}\label{section:related}
%%%%%%%%%%%%%%%%%%%%%%%%%%%%%%%%%%%%%%%%%%%%%%%%%%%%%%%%%%%%

\paragraph{Robustness and monitoring}
Robust (or quantitative) monitoring extends the binary question whether a log satisfies a specification by asking ``by how much'' the specification is satisfied.
The quantification of the distance between a signal and a signal temporal logic (STL) specification has been addressed in, \eg{} \cite{FP09,DM10,Donze10,DFM13,DMP17,JBGNN18} (or in a slightly different setting in~\cite{ALFS11}).
The distance can be understood in terms of space (``signals'') or time.
In~\cite{ABD18}, the distance also copes for reordering of events.
In~\cite{BFMU17}, the \emph{robust pattern matching problem} is considered over signal regular expressions, by quantifying the distance between the signal regular expression specification and the \emph{segments} of the signal.
For piecewise-constant and piecewise-linear signals, the problem can be effectively solved using a finite union of convex polyhedra.
% 	\mw{It looks the result is not zones but convex polyhedra if the signal is piecewise-linear. (It is already fixed)}\ea{thanks}
%
While our framework does not fit in robust monitoring, we can simulate both the robustness \wrt{} time (using timing parameters) and \wrt{} data, \eg{} signal values (using data parameters).

\paragraph{Monitoring with data}

The tool \textsc{MarQ}~\cite{RCR15} performs monitoring using Quantified Event Automata (QEA)~\cite{BFHRR12}.
This approach and ours share the automata-based framework, the ability to express some first-order properties using ``events containing data'' (which we encode using local variables associated with actions), and data may be quantified.
However, \cite{RCR15} does not seem to natively support specification parametric in time; in addition, \cite{RCR15} does not perform complete (``symbolic'') parameters synthesis, but outputs the violating entries of the log.
%\ea{we need to be careful, as this work is very close to ours}%
%\mw{I think they can also write timing parameter because it looks they handle timestamp as a usual rational data. However, they do not synthesize the parameter valuations, at least as we expect (I tried MarQ a bit.). It looks this parameter synthesis (\ie{}, symbolic monitoring, not a (non-symbolic) monitoring) is our main strength.}

The metric first order temporal logic (MFOTL) allows for a high expressiveness by allowing universal and existential quantification over data---which can be seen as a way to express parameters.
A monitoring algorithm is presented for a safety fragment of MFOTL in~\cite{BKMZ15}.
Aggregation operators are added in~\cite{BKMZ15FMSD}, allowing to compute \emph{sums} or \emph{maximums} over data.
A fragment of this logics is implemented in \MonPoly{}~\cite{BKZ17}.
While these works are highly expressive, they do not natively consider timing parameters; in addition, \MonPoly{} does not output symbolic answers, \ie{} symbolic conditions on the parameters to ensure validity of the formula.\reviewer{2}{In line 111 is mentioned timing parameters, but it is, I believe, not quite clear at this point what these are.}\todo{Thanks for your comment. We will make it clear by explaining the timing and data parameters around Example1.}

In~\cite{HPU17}, binary decision diagrams (BDDs) are used \LongVersion{in order }to symbolically represent the observed data in QTL.
This can be seen as monitoring data against a parametric specification, with a symbolic internal encoding\LongVersion{ (the BDDs of~\cite{HPU17,HP18} work efficiently for comparing whether a variable is equal or not equal to another, but not for comparing whether a variable is smaller than another one---which suits strings better than rationals)}.
However, their implementation \DejaVu{} only outputs \emph{concrete} answers.
In contrast, we are able to provide symbolic answers (both in timing and data parameters), \eg{} in the form of union of polyhedra for rationals, and unions of string constraints using equalities ($=$) and inequalities ($\neq$).
% 	\cite{HPU17} can be seen as one step forward in terms of symbolic representation thanks to~BDDs; however, .

% \todo{\cite{RCR15} with theory from~\cite{Reger14}}

\todo{RiTHM2, STePr}

\paragraph{Freeze operator}
In~\cite{BDSV14}, \LongVersion{the }STL \LongVersion{logic }is extended with a freeze operator that can ``remember'' the value of a signal, to compare it to a later value of the same signal.
This logic STL$^*$ can express properties such as ``In the initial 10 seconds, x copies the values of y within a delay of 4 seconds'':
\(\mathbf{G}_{[0,10]} \freeze ( \mathbf{G}_{[0, 4]} y^\freeze = x )\).
While the setting is somehow different (STL$^*$ operates over signals while we operate over timed data words), the requirements such as the one above can easily be encoded in our framework.
In addition, we are able to \emph{synthesize} the delay within which the values are always copied, as in \cref{example:copy}.
In contrast, it is not possible to determine using STL$^*$ which variables and which delays\LongVersion{ satisfy or} violate the specification.

%\ih{perhaps nice to \emph{fix} a leading specification}\ea{more or less done}

\paragraph{Monitoring with parameters}
In~\cite{ADMN11}, a log in the form of a dense-time real-valued signal is tested against a parameterized extension of STL, where parameters can be used to model uncertainty both in signal values and in timing values.
The output comes in the form of a subset of the parameters space for which the formula holds on the log.
In~\cite{BFM18}, the focus is only on signal parameters, with an improved efficiency by reusing techniques from the \emph{robust} monitoring.
% Our work can be seen as a generalization of~\cite{ADMN11,BFM18} in the automata-based framework instead of signals.\ea{OK?}\mw{I think the comparison is automata vs. temporal logic or events/timed words vs. signals, not automata vs. signals. For the other part, it thinks great.}
Whereas \cite{ADMN11,BFM18} fit in the framework of signals and temporal logics while we fit in words and automata,
	our work shares similarities with~\cite{ADMN11,BFM18} in the sense that we can express data parameters;
	in addition, \cite{BFM18} is able as in our work to exhibit the segment of the log associated with the parameters valuations for which the specification holds.
A main difference however is that we can use memory and aggregation, thanks to arithmetic on variables.
% 	A main difference however is that we also consider timing parameters; in addition, our data parameters may be more general than~\cite{BFM18} in the sense that we can work with arbitrary infinite-domain data, not only signals.\ea{new version; what do you think?}

In~\cite{FR08}, the %orthogonal
	problem of \emph{inferring} temporal logic formulae with constraints that hold in a given numerical data time series is addressed.
\LongVersion{The method is applied to biological systems.}\ea{I integrate this part in ``monitoring with parameters'' but it does not strictly speaking fit here.}

\paragraph{Timed pattern matching}
A recent line of work is that of timed pattern matching, that takes as input a log and a specification, and decides \emph{where} in the log the specification is satisfied or violated.
On the one hand, a line of works considers signals, with specifications either in the form of timed regular expressions~\cite{UFAM14,UFAM16,Ulus17,BFNMA18},
	or a temporal logic~\cite{UM18}%\mw{The specification language in~\cite{BFNMA18} is actually TREs, though the algorithm is automata-based. I also found a paper using temporal logic~\cite{UM18}}
.
On the other hand, a line of works considers timed words, with specifications in the form of timed automata~\cite{WHS17,AHW18}.
We will see that our work can also encode parametric timed pattern matching.
Therefore, our work can be seen as a two-dimensional extension of both lines of works: first, we add timing parameters (\LongVersion{note that~}\cite{AHW18} also considers similar timing parameters) and, second, we add data---themselves extended with parameters.
That is, coming back to \cref{example:copy}, \cite{UFAM14,UFAM16,Ulus17,WHS17} could only infer the segments of the log for which the property is violated for a given (fixed) variable and a given (fixed) timing parameter; while \cite{AHW18} could infer both the segments of the log and the timing parameter valuations, but not which variable violates the specification.
% 	\ea{new}
% 	\mw{Maybe it is not clear why our work is an extension of (parametric) timed pattern matching. Though I do not think it is an important point, we can add a remark if we need.}

\paragraph{Summary}
We compare related works \LongVersion{with our work }in \cref{table:comparisons}.
%%% TIMED ANSWER
% ``Timed answer'' denotes the ability to exhibit the log segments along which a specification is satisfied/violated (in contrast to a Boolean answer or a simple counter-example exhibition).
%
%%% TIMING PARAMETERS
``Timing parameters'' denote the ability to synthesize unknown constants used in timing constraints (\eg{} modalities intervals, or clock constraints).
``\cellUnclear{}'' denotes works not natively supporting this, although it might be encoded.\ea{to reread?}
%%% DATA
The term ``Data'' refers to the ability to manage logs over infinite domains (apart from timestamps).
	For example, the log in \cref{figure:examples:copy:log} features, beyond timestamps, both string (variable name) and rationals (value).
	Also, works based on real-valued signals are naturally able to manage (at least one type of) data.
%
%%% DATA PARAMETERS
``Parametric data'' refer to the ability to express formulas where data (including signal values) are compared to 
(quantified or unquantified) variables or unknown parameters; for example, in the log in \cref{figure:examples:copy:log}, an example of property parametric in data is to synthesize the parameters for which the difference of values between two consecutive updates of variable~$\styledparam{px}$ is always below~$\styledparam{pv}$, where $\styledparam{px}$ is a string parameter and $\styledparam{pv}$ a rational-valued parameter.
%\reviewer{2}{In Table 1 it is indicated that DejaVU supports parametric data. If I have understood
%this concept correctly, DejaVu does not support this concept. Can you explain? If it just means the ability to quantify over variables so: yes.
%}\todo{You are right. Here, ``parametric data'' means the ability to express the specification using (quantified or unquantified) variables, which is supported by DejaVu. We will make this point clearer.}
%
%%% MEMORY
``Memory'' is the ability to remember \emph{past} data; this can be achieved using \eg{} the freeze operator of STL$^*$, or variables (\eg{} in \cite{RCR15,BKMZ15,HPU17}).
%
%%% AGGREGATION
``Aggregation'' is the ability to aggregate data using operators such as sum or maximum; this allows to express properties such as ``A user must not withdraw more than \$10,000 within a 31 day period''~\cite{BKMZ15FMSD}.
This can be supported using dedicated aggregation operators~\cite{BKMZ15FMSD} or using variables (\cite{RCR15}, and our work).
%
%%% COMPLETE PARAMETER IDENTIFICATION
``Complete parameter identification'' denotes the \emph{synthesis} of the set of parameters that satisfy or violate the property.
Here, ``\cellNA{}'' denotes the absence of parameter~\cite{BDSV14}, or when parameters are used in a way (existentially or universally quantified) such as the identification is not explicit (instead, the position of the log where the property is violated is returned~\cite{HPU17}).
In contrast, we return in a \emph{symbolic} manner (as in~\cite{ADMN11,AHW18}) the exact set of (data and timing) parameters for which a property is satisfied.
``\cellYesNo{}'' denotes ``yes'' in the theory paper, but not in the \LongVersion{associated }tool.
% For example, returning only the position in the log when the property is violated is \emph{not} a complete parameter identification.
% \ea{Symbolic data can be misleading, as \cite{HPU17} is more symbolic than us}

% \ih{explain very well the table ``if we get this answer, this means…, if we write no, then…''}

% \ea{tentative table in \cref{table:comparisons}; should we add timed pattern matching (\eg{} \cite{UFAM14,WHS17}) or do we consider that \cite{AHW18} subsumes these works?}

%%%%%%%%%%%%%%%%%%%%%%%%%%%%%%%%%%%%%%%%%%%%%%%%%%%%%%%%%%%%
%%%%%%%%%%%%%%%%%%%%%%%%%%%%%%%%%%%%%%%%%%%%%%%%%%%%%%%%%%%%
\section{Preliminaries}\label{section:preliminaries}
%%%%%%%%%%%%%%%%%%%%%%%%%%%%%%%%%%%%%%%%%%%%%%%%%%%%%%%%%%%%
%%%%%%%%%%%%%%%%%%%%%%%%%%%%%%%%%%%%%%%%%%%%%%%%%%%%%%%%%%%%

%%%%%%%%%%%%%%%%%%%%%%%%%%%%%%%%%%%%%%%%%%%%%%%%%%%%%%%%%%%%
\LongVersion{\subsection{Clocks, timing parameters and timed guards}}
\ShortVersion{\subsubsection{Clocks, timing parameters and timed guards}}
%%%%%%%%%%%%%%%%%%%%%%%%%%%%%%%%%%%%%%%%%%%%%%%%%%%%%%%%%%%%

%Let $\interval(\grandn)$ denote the set of non-necessarily closed intervals on~$\grandn$, \ie{} of the form $[a,b]$, $(a,b]$, $[a,b)$ or $(a,b)$ where $a,b\in \grandn$ and $a \leq b$.

We assume a set~$\Clock = \{ \clock_1, \dots, \clock_\ClockCard \} $ of \emph{clocks}, \ie{} real-valued variables that evolve at the same rate.
A \emph{clock valuation} is\LongVersion{ a function}
$\clockval : \Clock \rightarrow \Rgeqzero$.
We write $\ClocksZero$ for the clock valuation assigning $0$ to all clocks.
Given $d \in \Rgeqzero$, $\clockval + d$ \ShortVersion{is}\LongVersion{denotes the valuation} \st{} $(\clockval + d)(\clock) = \clockval(\clock) + d$, for all $\clock \in \Clock$.
Given $\resets \subseteq \Clock$, we define the \emph{reset} of a valuation~$\clockval$, denoted by $\reset{\clockval}{\resets}$, as follows: $\reset{\clockval}{\resets}(\clock) = 0$ if $\clock \in \resets$, and $\reset{\clockval}{\resets}(\clock)=\clockval(\clock)$ otherwise.

We assume a set~$\TParam = \{ \tparam_1, \dots, \tparam_\TParamCard \} $ of \emph{timing parameters}\LongVersion{, \ie{} unknown timing constants}.
A \emph{timing parameter valuation} \LongVersion{$\tpval$ }is\LongVersion{ a function}
$\tpval : \TParam \rightarrow \grandqplus$.\LongVersion{\footnote{%
	We choose $\grandqplus$ by consistency with most of the PTA literature, but also because, for classical PTAs, choosing~$\Rgeqzero$ leads to undecidability~\cite{Miller00}.
}}
We assume ${\compOp} \in \{<, \leq, =, \geq, >\}$.
A \emph{timed guard}~$\tguard$ is a constraint over $\Clock \cup \TParam$ defined by a conjunction of inequalities of the form $\clock \compOp d$,
or $\clock \compOp \tparam$ with $d \in \grandn$ and $\tparam \in \TParam$.
Given~$\tguard$, we write~$\clockval \models \tpval(\tguard)$ if %$\valuate{\valuate{\tguard}{\tpval}}{\clockval}$
the expression obtained by replacing each~$\clock$ with~$\clockval(\clock)$ and each~$\tparam$ with~$\tpval(\tparam)$ in~$\tguard$ evaluates to true.

% A linear term over $\Clock \cup \TParam$ is of the form $\sum_{1 \leq i \leq \ClockCard} \alpha_i \clock_i + \sum_{1 \leq j \leq \TParamCard} \beta_j \tparam_j + d$, with
% 	$\clock_i \in \Clock$,
% 	$\tparam_j \in \TParam$,
% 	and
% 	$\alpha_i, \beta_j, d \in \grandz$.
%
% A \emph{constraint}~$\C$ (\ie{} a convex polyhedron) over $\Clock \cup \TParam$ is a conjunction of inequalities of the form $\lterm \compOp 0$, where $\lterm$ is a linear term.
%
% Given a set~$\TParam$ of parameters, we denote by $\project{\C}{\TParam}$ the projection of~$\C$ onto~$\TParam$, \ie{} obtained by eliminating the variables not in $\TParam$ (\eg{} using Fourier-Motzkin\LongVersion{~\cite{Schrijver86}}).
%\ea{removed defs of constraints and projection (so far)}
%\ea{need to add projection for Masaki}\mw{This projection is unnecessary and it is covered.}

%%%%%%%%%%%%%%%%%%%%%%%%%%%%%%%%%%%%%%%%%%%%%%%%%%%%%%%%%%%%
\ShortVersion{\subsubsection{Variables, data parameters and data guards}}
\LongVersion{\subsection{Variables, data parameters and data guards}}
%%%%%%%%%%%%%%%%%%%%%%%%%%%%%%%%%%%%%%%%%%%%%%%%%%%%%%%%%%%%

For sake of simplicity, we assume a \emph{single} infinite domain~$\Domain$ for data.
The formalism defined in \cref{section:formalism} can be extended in a straightforward manner to different domains for different variables (and our implementation \LongVersion{\masakiTool{} }does allow for different types).
The case of \emph{finite} data domain is immediate too.
We \LongVersion{however }define this formalism in an \emph{abstract} manner, so as to allow a sort of parameterized domain.

We assume a set~$\Variables = \{ \dvar_1, \dots, \dvar_\VariablesCard \} $ of \emph{variables} valued over~$\Domain$.
These variables are internal variables, that allow an high expressive power in our framework, as they can be compared or updated to other variables or parameters.
We also assume a set~$\LVariables = \{ \lvar_1, \dots, \lvar_\LVariablesCard \} $ of \emph{local variables} valued over~$\Domain$.
These variables will only be used locally along a transition in the ``argument'' of the action (\eg{} $\stylelvar{x}$ and $\stylelvar{v}$ in $\styleact{upate}(\stylelvar{x}, \stylelvar{v})$), and in the associated guard and (right-hand part of) updates.
We assume a set~$\VParam = \{ \vparam_1, \dots, \vparam_\VParamCard \} $ of \emph{data parameters}, \ie{} unknown variable constants.

A \emph{\datatype{}} $(\Domain, \DomainGuards , \DomainUpdates)$ is made of
\begin{ienumeration}
	\item an infinite domain $\Domain$,
	\item a set of admissible Boolean expressions~$\DomainGuards$ (that may rely on $\Variables$, $\LVariables$ and~$\VParam$), which will define the type of guards over variables in our subsequent automata,
	and
	\item a domain for updates $\DomainUpdates$ (that may rely on $\Variables$, $\LVariables$ and~$\VParam$), which will define the type of updates of variables in our subsequent automata.
\end{ienumeration}

%----------------------------------------------------------
\begin{example}
 \label{example:data-types}
	As a first example, let us define the \datatype{} for rationals.
	We have $\Domain = \grandq$.
	Let us define Boolean expressions.
	A \emph{rational comparison} is a constraint over $\Variables \cup \LVariables \cup \VParam$ defined by a conjunction of inequalities of the form
		$\dvar \compOp d$,
		$\dvar \compOp \dvar'$,
		or $\dvar \compOp \vparam$
		with $\dvar, \dvar' \in \Variables \cup \LVariables$, $d \in \grandq$ and $\vparam \in \VParam$.
	$\DomainGuards$ is the set of all rational comparisons over $\Variables \cup \LVariables \cup \VParam$.
	Let us then define updates.
	First, a linear arithmetic expression over $\Variables \cup \LVariables \cup \VParam$ is $\sum_i \alpha_i \dvar_i + \beta$, where $\dvar_i \in \Variables \cup \LVariables \cup \VParam$ and $\alpha_i, \beta \in \grandq$.
	Let $\ArExp(\Variables \cup \LVariables \cup \VParam)$ denote the set of arithmetic expressions over $\Variables$, $\LVariables$ and~$\VParam$.
	We then have $\DomainUpdates = \ArExp(\Variables \cup \LVariables \cup \VParam)$.
	
	As a second example, let us define the \datatype{} for strings.
	We have $\Domain = \grands$, where $\grands$ denotes the set of all strings.
	A \emph{string comparison} is a constraint over $\Variables \cup \LVariables \cup \VParam$ defined by a conjunction of comparisons of the form
		$\dvar \compString s$,
		$\dvar \compString \dvar'$,
		or $\dvar \compString \vparam$
	with $\dvar, \dvar' \in \Variables \cup \LVariables$, $s \in \grands$, $\vparam \in \VParam$ and ${\compString} \in \{ =, \neq \}$.
	$\DomainGuards$ is the set of all string comparisons over $\Variables \cup \LVariables \cup \VParam$.
	$\DomainUpdates = \Variables \cup \LVariables \cup \grands$, \ie{} a string variable can be assigned another string variable, or a concrete string.
% 		\ea{note that I don't allow assignment variable := parameter; is it necessary?}\mw{I think it is unnecessary. I do not think it is essential and it in not used in any examples in the paper.}
\end{example}
%----------------------------------------------------------

A \emph{variable valuation} is\LongVersion{ a function} $\varval : \Variables \rightarrow \Domain$.
A \emph{local variable valuation} is a partial function $\lvarval : \LVariables \partfun \Domain$.
A \emph{data parameter valuation} \LongVersion{$\dpval$ }is\LongVersion{ a function}
$\dpval : \VParam \rightarrow \Domain$.
% A comparison is a Boolean expression of the form \todo{} where $ {\datarel} \in \DataRels $\ea{argh, I'm stuck: how to express the fact that we may want coefficients to multiply rational-valued data, but such coefficients do not exist for strings…?}.
% A \emph{data guard} $\dguard$ is \todo{}
Given a data guard $\dguard \in \DomainGuards$,
a variable valuation~$\varval$,
a local variable valuation~$\lvarval$ defined for the local variables in $\dguard$,
and
a data parameter valuation~$\dpval$,
we write $(\varval, \lvarval) \models \valuate{\dguard}{\dpval}$
if
the expression obtained by replacing within $\dguard$ all occurrences of each data parameter~$\vparam_i$ by~$\dpval(\vparam_i)$ and all occurrences of each variable~$\dvar_j$ (resp.\ local variable $\lvar_k$) with its concrete valuation $\varval(\dvar_j)$ (resp.~$\lvarval(\lvar_k)$)) evaluates to true.

% \ea{define operations: assignments, …}

% Let \todo{} denote the set of \todo{} expressions over $\Variables$ and~$\VParam$.
A parametric data update is a partial function
	$\PDU : \Variables \partfun \DomainUpdates$.
	That is, we can assign to a variable an expression over data parameters and other variables, according to the \datatype{}.
Given a parametric data update~$\PDU$, a variable valuation~$\varval$, a local variable valuation~$\lvarval$ (defined for all local variables appearing in~$\PDU$), and a data parameter valuation~$\dpval$,
	we define $\DPValuate{\PDU}{\dpval}{\varval}{\lvarval} : \Variables \rightarrow \Domain$ as\LongVersion{ follows}:
\[\DPValuate{\PDU}{\dpval}{\varval}{\lvarval}(\dvar) = 
	\begin{cases} 
		\varval(\dvar) & \text{if $\PDU(\dvar) $ is undefined} \\
		\valuate{\valuate{\valuate{\PDU(\dvar)}{\dpval}}{\varval}}{\lvarval} & \text{otherwise} \\
	\end{cases}
\]
where $\valuate{\valuate{\valuate{\PDU(\dvar)}{\dpval}}{\varval}}{\lvarval}$ denotes the replacement within the update expression $\PDU(\dvar)$ of
	all occurrences of each data parameter~$\vparam_i$ by~$\dpval(\vparam_i)$,
	and
	all occurrences of each variable~$\dvar_j$ (resp.\ local variable $\lvar_k$) with its concrete valuation $\varval(\dvar_j)$ (resp.~$\lvarval(\lvar_k)$).
Observe that this replacement gives a value in~$\Domain$, therefore the result of $\DPValuate{\PDU}{\dpval}{\varval}{\lvarval}$ is indeed a data parameter valuation $\Variables \rightarrow \Domain$.
That is, $\DPValuate{\PDU}{\dpval}{\varval}{\lvarval}$ computes the new (non-parametric) variable valuation obtained after applying to~$\varval$ the partial function $\PDU$ valuated with~$\dpval$.

%----------------------------------------------------------
\begin{example}
	Consider the \datatype{} for rationals, the variables set $\{ \dvar_1, \dvar_2 \}$, the local variables set $\{ \lvar_1, \lvar_2 \}$ and the parameters set $\{ \vparam_1 \}$.
	Let $\varval$ be the variable valuation such that $\varval(\dvar_1) = 1$ and $\varval(\dvar_2) = 2$,
	and
	$\lvarval$ be the local variable valuation such that $\lvarval(\lvar_1) = 2$ and $\lvarval(\lvar_2)$ is not defined.
	Let $\dpval$ be the data parameter valuation such that $\dpval(\vparam_1) = 1$.
	Consider the parametric data update function~$\PDU$ such that $\PDU(\dvar_1) = 2 \times \dvar_1 + \dvar_2 - \lvar_1  + \vparam_1$, and $\PDU(\dvar_2)$ is undefined.
	Then the result of $\DPValuate{\PDU}{\dpval}{\varval}{\lvarval}$ is $\varval'$ such that $\varval'(\dvar_1) = 2 \times \varval(\dvar_1) + \varval(\dvar_2) - \lvarval(\lvar_1) + \dpval(\vparam_1) = 3$ and $\varval'(\dvar_2) = 2$.
% 		\ea{does it look ok? (I updated)}\mw{it looks good.}
\end{example}
%----------------------------------------------------------

%%%%%%%%%%%%%%%%%%%%%%%%%%%%%%%%%%%%%%%%%%%%%%%%%%%%%%%%%%%%
%%%%%%%%%%%%%%%%%%%%%%%%%%%%%%%%%%%%%%%%%%%%%%%%%%%%%%%%%%%%
\section{Parametric timed data automata}\label{section:formalism}
%%%%%%%%%%%%%%%%%%%%%%%%%%%%%%%%%%%%%%%%%%%%%%%%%%%%%%%%%%%%
%%%%%%%%%%%%%%%%%%%%%%%%%%%%%%%%%%%%%%%%%%%%%%%%%%%%%%%%%%%%

We introduce here Parametric timed data automata (PTDAs).
They can be seen as an extension of parametric timed automata~\cite{AHV93} (that extend timed automata~\cite{AD94} with parameters in place of integer constants) with unbounded data variables and parametric variables.
PTDAs can also be seen as an extension of some extensions of timed automata with data (see \eg{} \cite{BER94,Dang03,Quaas15}), that we again extend with both data parameters and timing parameters.\ea{more extensions to mention…?}
Or as an extension of quantified event automata~\cite{BFHRR12} with explicit time representation using clocks, and further augmented with timing parameters.
%  extend timed automata with parameters within guards %and invariants
PTDAs feature both timed guards and data guards;
we summarize the various variables and parameters types together with their notations in \cref{table:variables}.

%----------------------------------------------------------
\begin{table}[tb]
	\centering
	\LongVersion{\footnotesize}
	\ShortVersion{\scriptsize}

	\caption{Variables, parameters and valuations used in guards}
	\begin{tabular}{ l || c | c || c | c | c |}
% 		\hline
		& \multicolumn{2}{ c ||}{timed guards} & \multicolumn{3}{ c |}{data guards}
		\\ \hline
		& clock & timing parameter & (data) variable & local variable & data parameter
		\\ \hline
		Variable & $\clock$ & $\tparam$ & $\dvar$ & $\lvar$ & $\vparam$
		\\ \hline
		Valuation & $\clockval$ & $\tpval$ & $\varval$ & $\lvarval$ & $\dpval$
		\\ \hline
	\end{tabular}

	\label{table:variables}
\end{table}
%----------------------------------------------------------

\LongVersion{%
%%%%%%%%%%%%%%%%%%%%%%%%%%%%%%%%%%%%%%%%%%%%%%%%%%%%%%%%%%%%
\subsection{Syntax}
%%%%%%%%%%%%%%%%%%%%%%%%%%%%%%%%%%%%%%%%%%%%%%%%%%%%%%%%%%%%
}

We will associate local variables with actions (which can be see as \emph{predicates}).
Let $\ActDomain : \Actions \rightarrow 2^\LVariables$ denote the set of local variables associated with each action.
Let $\DomainVar(\dguard)$ (resp.~$\DomainVar(\PDU)$) denote the set of variables occurring in~$\dguard$ (resp.~$\PDU$).

%----------------------------------------------------------
\begin{definition}[PTDA]\label{def:PTDA}
	Given a \datatype{}~$(\Domain, \DomainGuards , \DomainUpdates)$,
	a parametric timed data automaton (PTDA) $\A$ over this \datatype{} is a tuple \mbox{$\A = (\Actions, \Loc, \locinit, \LocFinal, \Clock, \TParam, \Variables, \LVariables, \varvalo, \VParam, %\invariant,
		\Edges)$}, where:
% 			\todo{do we allow invariants?!}\mw{I did not allow invariants in my implementation. I suggest we do NOT have invariants.}
	\begin{enumerate}
		\item $\Actions$ is a finite set of actions,
		\item $\Loc$ is a finite set of locations,
		\LongVersion{\item }$\locinit \in \Loc$ is the initial location,
		\item $\LocFinal \subseteq \Loc$ is the set of accepting locations,
		\item $\Clock$ is a finite set of clocks,
		\item $\TParam$ is a finite set of timing parameters,
		\item $\Variables$ (resp.~$\LVariables$) is a finite set of variables (resp.\ local variables) over~$\Domain$,
		\item $\varvalo$ is the initial variable valuation,
		\item $\VParam$ is a finite set of data parameters,
%		\item $\invariant$ is the invariant, assigning to every $\loc\in \Loc$ a guard $\invariant(\loc)$,\mw{In my algorithm, implementation, and examples, invariant is not considered. Do you mind removing invariant from the definition?}\ea{OK}
		\item $\Edges$ is a finite set of edges  $\edge = (\loc, \tguard, \dguard, \action, \resets, \PDU, \loc')$
		where
		\begin{ienumeration}
			\item $\loc, \loc'\in \Loc$ are the source and target locations,
			\item $\tguard$ is a timed guard,
			\item $\dguard \in \DomainGuards$ is a data guard such as $\DomainVar(\dguard) \cap \LVariables \subseteq \ActDomain(\action)$,
			\item $\action \in \Actions$,
			\item $\resets\subseteq \Clock$ is a set of clocks to be reset, and
			\item $\PDU : \Variables \partfun \DomainUpdates$ is the parametric data update function such that $\DomainVar(\PDU) \cap \LVariables \subseteq \ActDomain(\action)$.
		\end{ienumeration}
	\end{enumerate}
\end{definition}
%----------------------------------------------------------

The domain conditions on $\dguard$ and $\PDU$ ensure that the local variables used in the guard (resp.~update) are only those in the action signature $\ActDomain(\action)$.

%----------------------------------------------------------
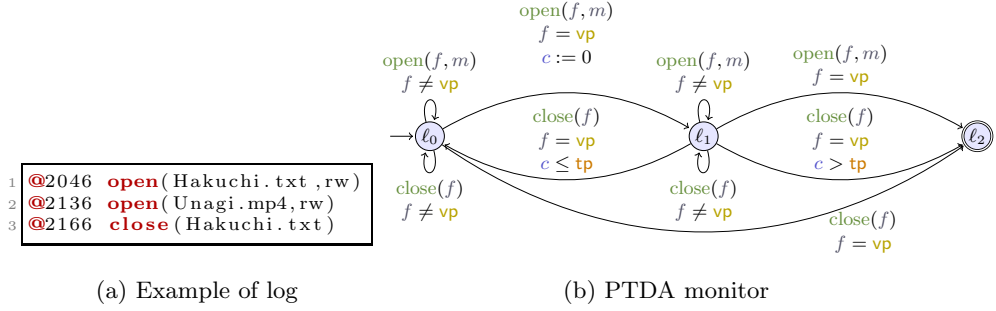
\begin{figure*}[t]
	%----------------------------------------------------------
	\begin{subfigure}[b]{.359\textwidth}
		\begin{lstlisting}[style=log]
@2046 open(Hakuchi.txt,rw)
@2136 open(Unagi.mp4,rw)
@2166 close(Hakuchi.txt)
\end{lstlisting}
		\caption{Example of log}
		\label{code:log}
	\end{subfigure}
	%----------------------------------------------------------
	%
	\hfill
	\begin{subfigure}[b]{.64\textwidth}
	\centering
		\footnotesize
		\scalebox{.8}{
		\begin{tikzpicture}[shorten >=1pt,node distance=4.5cm,on grid,auto]
		%% states
		\node[location,initial] (s_0) {$\loc_0$};
		\node[location] (s_1) [right of=s_0] {$\loc_1$};
		\node[location,accepting] (s_2) [right of=s_1] {$\loc_2$};

% 		%% edges
		\path[->] 
		(s_0) edge [loop above] node[align=center] {$\styleact{open}(\stylelvar{f}, \stylelvar{m})$ \\ $\stylelvar{f} \neq \styledparam{\vparam}$} (s_0)
		
		(s_0) edge [loop below] node[align=center] {$\styleact{close}(\stylelvar{f})$ \\ $\stylelvar{f} \neq \styledparam{\vparam}$} (s_0)
		
		(s_0) edge [bend left, above] node[align=center] {$\styleact{open}(\stylelvar{f}, \stylelvar{m})$\\
										$\stylelvar{f} = \styledparam{\vparam}$ \\
                                          $\styleclock{\clock} := 0$\\
                                         } (s_1)

		(s_1) edge [loop above] node[align=center] {$\styleact{open}(\stylelvar{f}, \stylelvar{m})$ \\ $\stylelvar{f} \neq \styledparam{\vparam}$} (s_1)
		
		(s_1) edge [loop below] node[align=center] {$\styleact{close}(\stylelvar{f})$ \\ $\stylelvar{f} \neq \styledparam{\vparam}$} (s_1)

		(s_1) edge [bend right, above] node[align=center] {$\styleact{close}(\stylelvar{f})$ \\ $\stylelvar{f} = \styledparam{\vparam}$ \\ $\styleclock{\clock} > \styleparam{\tparam}$} (s_2)
		
		(s_1) edge [bend left, above] node[align=center] {$\styleact{open}(\stylelvar{f}, \stylelvar{m})$ \\ $\stylelvar{f} = \styledparam{\vparam}$ } (s_2)
		
		(s_1) edge [bend left, above] node[align=center] {$\styleact{close}(\stylelvar{f})$ \\ $\stylelvar{f} = \styledparam{\vparam}$ \\ $\styleclock{\clock} \leq \styleparam{\tparam}$} (s_0)
		
		(s_0) edge [bend angle=35, bend right] node[right, xshift=6em, align=center] {$\styleact{close}(\stylelvar{f})$ \\ $\stylelvar{f} = \styledparam{\vparam}$ } (s_2)
	;
		\end{tikzpicture}}
	\caption{PTDA monitor}
	\label{figure:example:files}
	\end{subfigure}
	
	\caption{Monitoring proper file opening and closing}
	\label{figure:example:file}
\end{figure*}
%----------------------------------------------------------

%----------------------------------------------------------
\begin{example}
% 	\ea{Masaki, once a case study is stabilized, please let me know, I'll use it for example here}
	Consider the PTDA in \cref{figure:example:files} over the \datatype{} for strings.
	We have
		$\Clock = \{ \clock\}$,
		$\TParam = \{ \tparam \}$,
		$\Variables = \emptyset$
		and
		$\LVariables = \{ f , m \}$.
		$\ActDomain(\styleact{open}) = \{ f, m \}$ while $\ActDomain(\styleact{close}) = \{ f \}$.
		\todo{replace the late close by a deadline? But need unobservable actions…}
		$\loc_2$ is the only accepting location, modeling the violation of the specification.
		
		This PTDA (freely inspired by a formula from~\cite{HPU17} further extended with timing parameters)\ea{technically, we are much more concise and more expressive} monitors the improper file opening and closing, \ie{} a file already open should not be open again, and a file that is open should not be closed too late.
		The data parameter $\vparam$ is used to \emph{symbolically} monitor a given file name, \ie{} we are interested in opening and closings of this file only, while other files are disregarded (specified using the self-loops in $\loc_0$ and~$\loc_1$ with data guard $f \neq \vparam$).
		Whenever $f$ is opened (transition from~$\loc_0$ to~$\loc_1$), a clock~$\clock$ is reset.
		Then, in~$\loc_1$, if $f$ is closed within $\tparam$ time units (timed guard ``$\clock \leq \tparam$''), then the system goes back to~$\loc_0$.
		However, if instead $f$ is opened again, this is an incorrect behavior and the system enters~$\loc_2$ via the upper transition.
		The same occurs if $f$ is closed more than $\tparam$ time units after opening.
\end{example}
%----------------------------------------------------------

Given a data parameter valuation~$\dpval$
and
a timing parameter valuation~$\tpval$,
we denote by $\valuate{\A}{\dtpval{\tpval}{\dpval}}$ the resulting \emph{timed data automaton (TDA)}, \ie{} the non-parametric structure where all occurrences of a parameter~$\vparam_i$ (resp.~$\tparam_j$) have been replaced by~$\dpval(\vparam_i)$ (resp.~$\tpval(\tparam_j)$).
\LongVersion{%

}%
Note that, if $\Variables = \LVariables = \emptyset$, then $\A$ is a \emph{parametric timed automaton}~\cite{AHV93} and $\valuate{\A}{\dtpval{\tpval}{\dpval}}$ is a \emph{timed automaton}~\cite{AD94}.\LongVersion{\footnote{%
	We may need to multiply all timing constants in $\valuate{\A}{\dtpval{\tpval}{\dpval}}$ by the least common multiple of their denominators, so as to obtain an equivalent (integer-valued) TA\LongVersion{, as defined in~\cite{AD94}}.
}}

\LongVersion{
%%%%%%%%%%%%%%%%%%%%%%%%%%%%%%%%%%%%%%%%%%%%%%%%%%%%%%%%%%%%
\subsection{Semantics}
%%%%%%%%%%%%%%%%%%%%%%%%%%%%%%%%%%%%%%%%%%%%%%%%%%%%%%%%%%%%
}
\ShortVersion{
\medskip
}

We now equip our TDAs with a concrete semantics.

%----------------------------------------------------------
\begin{definition}[Semantics of a TDA]
	Given a PTDA \mbox{$\A = (\Actions, \Loc, \locinit, \LocFinal, \Clock, \TParam, \Variables, \LVariables, \varvalo, \VParam, %\invariant,
		\Edges)$} over a \datatype{} $(\Domain, \DomainGuards , \DomainUpdates)$,
	a data parameter valuation~$\dpval$
	and
	a timing parameter valuation~$\tpval$,
	the semantics of $\valuate{\A}{\dtpval{\tpval}{\dpval}}$ is given by the timed transition system (TTS) $(\States, \sinit, \flecheRel)$, with
	\begin{itemize}
		\item $\States = \Loc \times \Domain^\VariablesCard \times \Rgeqzero^\ClockCard$, %\{ (\loc, \clockval) \in \Loc \times \Rgeqzero^\ClockCard \mid \clockval \models \valuate{\invariant(\loc)}{\tpval} \}$, % \valuate{\valuate{\invariant(\loc)}{\tpval}}{\clockval} \text{ evaluates to true} 
		\LongVersion{\item} $\sinit = (\locinit, \varvalo, \ClocksZero) $,
		\item  $\flecheRel$ consists of the discrete and (continuous) delay transition relations:
		\begin{enumerate}
			\item discrete transitions: $(\loc, \varval, \clockval) \longueflecheRel{\edge, \lvarval} (\loc', \varval', \clockval')$, %with $\action \in \Sigma$,
				if
					%$(\loc, \varval, \clockval) , (\loc', \varval', \clockval') \in \States$, and
				there exist
					$\edge = (\loc, \tguard, \dguard, \action, \resets, \PDU, \loc') \in \Edges$
					and
					a local variable valuation~$\lvarval$ defined exactly for $\ActDomain(\action)$,
					such that
					$\clockval \models \tpval(\tguard$),
					$(\varval, \lvarval) \models \valuate{\dguard}{\dpval}$,
					$\clockval' = \reset{\clockval}{\resets}$,
				and
					$\varval' = \DPValuate{\PDU}{\dpval}{\varval}{\lvarval}$.
				% $\valuate{\valuate{\tguard}{\tpval}}{\clockval}$ evaluates to true.
			\item delay transitions: $(\loc, \varval, \clockval) \longueflecheRel{d} (\loc, \varval, \clockval + d)$, with $d \in \Rgeqzero$. %, if $\forall d' \in [0, d], (\loc, \clockval+d') \in \States$.
%                  \mw{It is just a matter of notation, but if we want to have a similar notation for both timing and data, $\varval \models \valuate{\dguard}{\dpval}$ and $\reset{\varval}{\valuate{\PDU}{\dpval}}$ might be also sensible.}\ea{good idea, fixed}
		\end{enumerate}
	\end{itemize}
\end{definition}

    Moreover we write
% 		$(\loc, \varval, \clockval)\longuefleche{(\edge, d)} (\loc', \varval', \clockval')$
		$\large((\loc, \varval, \clockval) , (\edge, \lvarval, d) , (\loc', \varval', \clockval')\large) \in \flecheRel$
		for a combination of a delay and discrete transition if
		% d,
		$\exists  \clockval'' :  (\loc, \varval, \clockval) \longueflecheRel{d} (\loc, \varval, \clockval'') \longueflecheRel{\edge, \lvarval} (\loc', \varval', \clockval')$.

Given a TDA~$\valuate{\A}{\dtpval{\tpval}{\dpval}}$ with concrete semantics $(\States, \sinit, \flecheRel)$, we refer to the states of~$\States$ as the \emph{concrete states} of~$\valuate{\A}{\dtpval{\tpval}{\dpval}}$.
A \emph{run} of~$\valuate{\A}{\dtpval{\tpval}{\dpval}}$ is an alternating sequence of concrete states of $\valuate{\A}{\dtpval{\tpval}{\dpval}}$ and triples of edges, local variable valuations and delays, starting from the initial state $\sinit$ of the form
$(\loc_0, \varval_0, \clockval_0), (\edge_0, \lvarval, d_0), (\loc_1, \varval_1, \clockval_1), \cdots$
with
$i = 0, 1, \dots$, $\edge_i \in \Edges$, $d_i \in \Rgeqzero$ and $\large((\loc_i, \varval_i, \clockval_i) , (\edge_i , \lvarval_i, d_i) , (\loc_{i+1}, \varval_{i+1}, \clockval_{i+1})\large) \in \flecheRel$.
Given such a run, the associated \emph{timed data word} is $(\action_1, \tau_1, \lvarval_1), (\action_2, \tau_2, \lvarval_2), \cdots$,
	where $\action_i$ is the action of edge~$\edge_{i-1}$,
	$\lvarval_i$ is the local variable valuation associated with that transition,
	and
	$\tau_i = \sum_{0 \leq j \leq i-1} d_j$, for $i = 1, 2 \cdots$.%
% 		\mw{Data in the input ``timed word'' is missing. I suppose our ``timed word'' can be something like $(\action_1, \tau_1, u_1), (\action_2, \tau_2, u_2), \cdots$, where $u_i$ is the value of the data. ($u_i \colon \Variables \partfun \Domain$ ?)}\ea{almost agree, but the function should be total: we know the initial variable valuation, therefore all subsequent valuations are total functions}\ea{fixed, what do you think?}\mw{I expected $(\action_1, \tau_1, \lvarval_1), (\action_2, \tau_2, \lvarval_2), \cdots$ rather than $(\action_1, \tau_1, \varval_0, \lvarval_1), (\action_2, \tau_2, \varval_1, \lvarval_2), \cdots$.}\ea{true; more inline with the input words. I change accordingly;}%
	\LongVersion{\footnote{%
	The ``$-1$'' in indices comes from the fact that, following usual conventions in the literature, states are numbered starting from~0 while words are numbered from~1.
}}
For a timed data word $\word$ and a concrete state $(\loc,\varval,\clockval)$ of\LongVersion{ a TDA}~$\valuate{\A}{\dtpval{\tpval}{\dpval}}$, we write
$(\locinit, \varvalo, \ClocksZero) \xrightarrow{\word} (\loc,\varval,\clockval)$ in $\valuate{\A}{\dtpval{\tpval}{\dpval}}$ 
if $\word$ is associated with a run of $\valuate{\A}{\dtpval{\tpval}{\dpval}}$ of the form $(\locinit, \varvalo, \ClocksZero), \dots, (\loc_n,\varval_n,\clockval_n)$ with $(\loc_n,\varval_n,\clockval_n) = (\loc,\varval,\clockval)$.
For a timed data word $\word = (\action_1, \tau_1, \lvarval_1), (\action_2, \tau_2, \lvarval_2),\dots, (\action_n, \tau_n, \lvarval_n)$, we denote $|\word| = n$ and 
for any $i \in \{1,2,\dots,n\}$, we denote $\word(1,i) = (\action_1, \tau_1, \lvarval_1), (\action_2, \tau_2, \lvarval_2),\dots, (\action_i, \tau_i, \lvarval_i)$.

% In the future, we will usually only give the projection of~$\lvarval_i$ onto $\ActDomain(\action_i)$, as only the valuation of the subset of the local variables associated with action~$\action_i$ is of interest.
%
% BEGIN REACHABILITY: USELESS
% Given\LongVersion{ a state}~$\state=(\loc, \varval, \clockval)$, we say that $\state$ is reachable in~$\valuate{\A}{\dtpval{\tpval}{\dpval}}$ if $\state$ appears in a run of $\valuate{\A}{\dtpval{\tpval}{\dpval}}$.
% By extension, we say that $\loc$ is reachable; and by extension again, given a set~$\somelocs$ of locations, we say that $\somelocs$ is reachable if there exists $\loc \in \somelocs$ such that $\loc$ is reachable in~$\valuate{\A}{\dtpval{\tpval}{\dpval}}$.\ea{useful?!}\mw{Probably I will not use it.}
% END REACHABILITY: USELESS

A finite run is \emph{accepting} if its last state $(\loc, \varval, \clockval)$ is such that $\loc \in \LocFinal$.
The \emph{language} $\Lg(\valuate{\A}{\dtpval{\tpval}{\dpval}})$ is defined to be the set of timed data words associated with all accepting runs of~$\valuate{\A}{\dtpval{\tpval}{\dpval}}$.
% 	\ea{do we need the timed language…?}\mw{(Partially) Yes. The language $\Lg(\valuate{\A}{\dtpval{\tpval}{\dpval}})$, which is a set of timed data words, is used in the definition of symbolic monitoring.}\ea{why ``partially''? We need it, right?}\mw{Yes. We need it.}

%----------------------------------------------------------
\begin{example}
	Consider \LongVersion{again }the PTDA in \cref{figure:example:files} over the \datatype{} for strings.
	Let $\tpval(\tparam) = 100$ and $\dpval(\vparam) = \mathtt{Hakuchi.txt}$.
	An accepting run of the TDA~$\valuate{\A}{\dtpval{\tpval}{\dpval}}$ is:
	\\
	$(\loc_0, \emptyset, \clockval_0),
		(\edge_0, \lvarval_0, 2046),
		(\loc_1, \emptyset, \clockval_1),
		(\edge_1, \lvarval_1, 90),
		(\loc_1, \emptyset, \clockval_2)
		(\edge_2, \lvarval_2, 30),
		(\loc_2, \emptyset, \clockval_3)
	$,
	\\
	where $\emptyset$ denotes a variable valuation over an empty domain (recall that $\Variables = \emptyset$ in \cref{figure:example:files}),
	$\clockval_0(\clock) = 0$,
	$\clockval_1(\clock) = 0$,
	$\clockval_2(\clock) = 90$,
	$\clockval_3(\clock) = 120$,
	$\edge_0$ is the upper edge from~$\loc_0$ to~$\loc_1$,
	$\edge_1$ is the self-loop above~$\loc_1$,
	$\edge_2$ is the lower edge from~$\loc_1$ to~$\loc_2$,
	$\lvarval_0(f) = \lvarval_2(f) = \mathtt{Hakuchi.txt}$,
	$\lvarval_1(f) = \mathtt{Unagi.mp4}$,
	$\lvarval_0(m) = \lvarval_1(m) = \mathtt{rw}$,
	and $\lvarval_2(m)$ is undefined (because $\ActDomain(\styleact{close}) = \{ f \}$).
	
	The associated timed data word is
	$
	(\styleact{open}, 2046, \lvarval_0),
	(\styleact{open}, 2136, \lvarval_1),
	(\styleact{close}, 2166, \lvarval_2)$.
	
	Since each action is associated with a set of local variables, given an ordering on this set, it is possible to see a given action and a variable valuation as a predicate:
% 		\ea{not sure whether this naming is appropriate?}\mw{I think the naming is fine.}
	for example, assuming an ordering of $\LVariables$ such as $f$ precedes~$m$, then $\styleact{open}$ with $\lvarval_0$ can be represented as $\styleact{open}(\mathtt{Hakuchi.txt}, \mathtt{rw})$.
	Using this convention, the log in \cref{code:log} corresponds exactly to this timed data word.
\end{example}
%----------------------------------------------------------

\begin{comment}

 %%%%%%%%%%%%%%%%%%%%%%%%%%%%%%%%%%%%%%%%%%%%%%%%%%%%%%%%%%%%
 %%%%%%%%%%%%%%%%%%%%%%%%%%%%%%%%%%%%%%%%%%%%%%%%%%%%%%%%%%%%
 \section*{Some interesting formulas}
 %%%%%%%%%%%%%%%%%%%%%%%%%%%%%%%%%%%%%%%%%%%%%%%%%%%%%%%%%%%%
 %%%%%%%%%%%%%%%%%%%%%%%%%%%%%%%%%%%%%%%%%%%%%%%%%%%%%%%%%%%%

 \ea{this section probably won't remain as such, but I use it to store interesting examples}

 \ea{\cite{HPU17}: $\forall f (close( f ) \Rightarrow \mathsf{Past} open( f ))$}

 \ea{\cite{HPU17}: $\forall f (close( f ) \Rightarrow \mathsf{Previous} (\neg close( f ) \mathsf{Since} open( f )))$}

 \ea{I converted both formulas to our formalism; not sure it's clearly simpler}

 \ea{\cite{BDSV14}: $G_{[0,60]} \big( ( F_{[0,10]} * [ G_{[1,10]} x^* \geq x + c]) \land (F_{[0,10]} * [G_{[1, 10]} x^* \leq x - c ]) \big)$}

 \ea{\cite{BDSV14}: $G_{[0,80 - (4 + \delta)]} * [G_{[4- \delta, 4+ \delta]} y^* = x]$}
\end{comment}

%%%%%%%%%%%%%%%%%%%%%%%%%%%%%%%%%%%%%%%%%%%%%%%%%%%%%%%%%%%%
%%%%%%%%%%%%%%%%%%%%%%%%%%%%%%%%%%%%%%%%%%%%%%%%%%%%%%%%%%%%
\section{Symbolic monitoring against PTDA specifications}\label{section:monitoring}
%%%%%%%%%%%%%%%%%%%%%%%%%%%%%%%%%%%%%%%%%%%%%%%%%%%%%%%%%%%%
%%%%%%%%%%%%%%%%%%%%%%%%%%%%%%%%%%%%%%%%%%%%%%%%%%%%%%%%%%%%

%In this section, we introduce the \emph{symbolic monitoring} problem and present an online algorithm to solve it.

In symbolic monitoring, in addition to the (observable) actions in $\Actions$, we employ \emph{unobservable} actions denoted by $\uAction$ and satisfying $\ActDomain(\uAction) = \emptyset$.
% 	\ea{not really anonymous!}\mw{maybe not really, but I suppose it is important to claim: 1) our problem can do something impossible for other problem settings/tools (e.g., MonPoly); and 2) our problem setting is actually a generalization of an existing problem setting (e.g., parametric timed pattern matching). So far, I do not have any better idea than comparing with \cite{AHW18}.}
We write $\ActionsWithU$ for $\Actions \disjointUnion \{\uAction\}$.
We let $\lvarval_{\uAction}$ be the local variable valuation such that $\lvarval_{\uAction}(\lvar)$ is undefined for any $\lvar \in \LVariables$.
For a timed data word $\word = (\action_1,\tau_1,\lvarval_1),(\action_2,\tau_2,\lvarval_2),\dots,(\action_n,\tau_n,\lvarval_n)$ over $\ActionsWithU$, the projection $\project{\word}{\Actions}$ is the timed data word over $\Actions$ obtained from~$\word$ by removing any triple $(\action_i,\tau_i,\lvarval_i)$ where $\action_i = \uAction$.
% For a PTDA~$\A$ over $\Actions_{\varepsilon}$, we denote $\LgU(\A) = \{\word \mid \exists \word' \in \Lg(\A).\, \project{\word'}{\Actions} = \word\}$.
An edge $\edge = (\loc, \tguard, \dguard, \action, \resets, \PDU, \loc') \in \Edges$ is \emph{unobservable} if $\action = \uAction$, and \emph{observable} otherwise.
The use of unobservable actions\LongVersion{ makes symbolic monitoring more general, and} allows us \LongVersion{in particular }to encode parametric timed pattern matching (see \cref{ss:PTPM}).

\LongVersion{
\begin{example}
 Let $\Actions = \{\styleact{wd}\}$, $\LVariables = \{a\}$, and $\ActDomain(\styleact{wd}) = \{a\}$.
 For a timed data word $\word = (\styleact{wd}, 0.1,\lvarval_1), (\styleact{wd}, 0.3, \lvarval_2), (\uAction, 0.7, \lvarval_3), (\styleact{wd}, 0.9, \lvarval_4)$ over $\ActionsWithU$, 
 $\project{\word}{\Actions}$ is the timed data word
 $\project{\word}{\Actions} = (\styleact{wd}, 0.1, \lvarval_1), (\styleact{wd}, 0.3, \lvarval_2),(\styleact{wd}, 0.9, \lvarval_4)$ over $\Actions$, where
 $\lvarval_1(a) = 10$, $\lvarval_2(a) = 10$, $\lvarval_3(a)$ is undefined, and $\lvarval_4(a) = 30$.%\ea{I don't know whether this example is really necessary? perhaps, if we need space, we can put in appendix or keep it for the long version?}\mw{I agree.}
\end{example}
}

We make the following assumption on the PTDAs in symbolic monitoring.

\begin{assumption}
 \label{assumption:no_unobservable_loop}
 The PTDA $\A$ does not contain any loop of unobservable edges.
\end{assumption}

\subsection{Problem definition}

Roughly speaking, given a PTDA $\A$ and a timed data word $\word$, 
the symbolic monitoring problem asks for the set of pairs $(\tpval, \dpval) \in \PVal \times \Domain^\VParam$ 
satisfying $\word(1,i) \in \valuate{\A}{\dtpval{\tpval}{\dpval}}$, 
where $\word(1,i)$ is a prefix of $\word$.
Since $\A$ also contains unobservable edges, we consider $\word'$ which is $\word$ augmented by unobservable actions.

\defProblem
	{Symbolic monitoring}
	{a PTDA~$\A$ over a \datatype{}~$(\Domain, \DomainGuards , \DomainUpdates)$ and actions $\ActionsWithU$, and a timed data word~$\word$ over $\Actions$}
	{compute all the pairs $(\tpval, \dpval)$ of timing and data parameter valuations such that there is a timed data word $\word'$ over $\ActionsWithU$ and $i\in\{1,2,\dots,|\word'|\}$ satisfying 
        $\project{\word'}{\Actions} = \word$ and $\word'(1,i) \in \Lg(\valuate{\A}{\dtpval{\tpval}{\dpval}})$.
	That is, it requires the \emph{validity domain}
        $\ValidityDomain{\word}{\A} = \{(\tpval,\dpval) \mid \exists \word' : i\in\{1,2,\dots,|\word'|\} , \project{\word'}{\Actions} = \word \text{ and } \word'(1,i) \in \Lg(\valuate{\A}{\dtpval{\tpval}{\dpval}}) \}$.%\mw{The term ``validity domain'' is taken from~\cite{ADMN11}.}
}

\begin{example}
 \label{example:symbolic_monitoring}
 Consider the PTDA $\A$ and the timed data word $\word$ shown in \cref{figure:example:copy}.
 The validity domain $\ValidityDomain{\word}{\A}$ is $\ValidityDomain{\word}{\A} = D_1 \cup D_2$, where
 {\small
 \begin{displaymath}
  D_1 = \big\{(\tpval, \dpval) \mid 0 \leq \tpval(\styleparam{tp}) \leq 2, \dpval(\styledparam{xp}) = \textrm{c} \big\}\text{ and }
  D_2 = \big\{(\tpval, \dpval) \mid 0 \leq \tpval(\styleparam{tp}) \leq 1, \dpval(\styledparam{xp}) = \textrm{a}\big\} \text{.}
 \end{displaymath}}%
 For $\word' = \word(1,3) \cdot (\uAction, \lvarval_{\uAction}, 2.9)$,
 we have $\word' \in \Lg(\valuate{\A}{\dtpval{\tpval}{\dpval}})$ and $\project{\word'}{\Actions} = \word(1,3)$, where
 $\tpval$ and $\dpval$ are such that
 $\tpval(\styleparam{tp}) = 1.8$ and $\dpval(\styledparam{xp}) = \textrm{c}$, and
 $\word(1,3) \cdot (\uAction, \lvarval_{\uAction}, 2.9)$ denotes the juxtaposition.
% Let $\A$ and $\word$ be the PTDA and timed data word in \cref{figure:running-example}, where the \datatype{} is that for rationals in \cref{example:data-types}.
% The validity domain $\ValidityDomain{\word}{\A}$ is represented by the union of the parameter valuations represented by the constraints in \cref{example:symbolic_monitoring:eq1,example:symbolic_monitoring:eq2,example:symbolic_monitoring:eq3}.
% The validity domain $\ValidityDomain{\word}{\A}$ shows the periods between two successive $\mathrm{wd}(a)$ satisfying $a > \vparam$.
% Using the data parameter $\vparam$, we observe the period of:
% \begin{ienumeration}
%  \item both $\mathrm{wd}(30)$ and $\mathrm{wd}(100)$ (\cref{example:symbolic_monitoring:eq1});
%  \item only $\mathrm{wd}(100)$ (\cref{example:symbolic_monitoring:eq2}); and
%  \item none of them (\cref{example:symbolic_monitoring:eq3}).  
% \end{ienumeration}
\reviewer{2}{It might be useful for the reader with a bit more explanation of Example 6. One could even imagine
such an explanation coming early in the paper. No need to do this, just a thought.}
\todo{Indeed! We will make Example1 more complete. Thanks!}
\end{example}

For the \datatype{}s in \cref{example:data-types}, the validity domain $\ValidityDomain{\word}{\A}$ can be represented by a constraint of finite size because the length $|\word|$ of the timed data word is finite.

%The parameter valuations $(\tpval, \dpval)$ represented by \cref{example:symbolic_monitoring:eq1} belong to $\ValidityDomain{\word}{\A}$ because
%$(a(5), 7), (a(10), 20), (\uAction,t) \in \Lg(\valuate{\A}{\dtpval{\tpval}{\dpval}})$,
%where $t \in (20,30)$; and
%the parameter valuations $(\tpval, \dpval)$ represented by \cref{example:symbolic_monitoring:eq2} belong to $\ValidityDomain{\word}{\A}$ because
%$(a(5), 7), (a(10), 20), (a(-5), 41), (\uAction,t) \in \Lg(\valuate{\A}{\dtpval{\tpval}{\dpval}})$,
%where $t \in (41,51)$.
%\todo{Probably, the notation will be changed.}

\subsection{Online algorithm}

Our algorithm is \emph{online} in the sense that it outputs $(\tpval, \dpval) \in \ValidityDomain{\word}{\A}$ as soon as its membership is witnessed, even before reading the whole timed data word $\word$.
%Precisely, after reading the $i$-th element $(\action_i, \tau_i)$ of the \todo{timed data word} $\word$, it immediately outputs all the pairs $(\tpval, \dpval)$ such that there is a timed data word $\word'$ over $\ActionsWithU$, $j\in\{1,2,\dots,|\word'|\}$ satisfying $\project{\word'}{\Actions} = \word$ $\word'(1,j) \in \Lg(\valuate{\A}{\dtpval{\tpval}{\dpval}})$, and if $a'_j \neq \uAction$, $\project{\word'(1,j)}{\Actions} = \word(1,i)$ and otherwise, $\project{\word'(1,j)}{\Actions} = \word(1,i-1)$.
%Our algorithm finds all the parameter valuations $(\tpval, \dpval) \in \ValidityDomain{\word}{\A}$ by a breadth-first search.

\LongVersion{\paragraph{Outline}}
Let $\word = (\action_1,\tau_1,\lvarval_1), (\action_2,\tau_2,\lvarval_2), \dots (\action_n,\tau_n,\lvarval_n)$ and $\A$ be the timed data word and PTDA given in symbolic monitoring, respectively.
Intuitively,
after reading $(\action_i,\tau_i,\lvarval_i)$, 
our algorithm symbolically computes
for all parameter valuations $(\tpval,\dpval) \in \PVal \times \Domain^{\VParam}$
the concrete states $(\loc,\clockval,\varval)$ satisfying
$(\locinit, \varvalo, \ClocksZero) \xrightarrow{\word(1,i)} (\loc, \varval, \clockval)$ in $\valuate{\A}{\dtpval{\tpval}{\dpval}}$.
Since $\A$ has unobservable edges as well as observable edges, we have to add unobservable actions before or after observable actions in $\word$.
By $\ObservableConf{i}$, 
we denote the configurations after reading $(\action_i,\tau_i,\lvarval_i)$ and no unobservable actions are appended after $(\action_i,\tau_i,\lvarval_i)$.
By $\UnobservableConf{i}$, 
we denote the configurations after reading $(\action_i,\tau_i,\lvarval_i)$ and at least one unobservable action is appended after $(\action_i,\tau_i,\lvarval_i)$.

\begin{definition}
 [$\ObservableConf{i}$, $\UnobservableConf{i}$]
 For 
 a PTDA~$\A$ over actions $\ActionsWithU$, 
 a timed data word $\word$ over $\Actions$, and
 $i \in \{0,1,\dots,|\word|\}$ (resp.~$i \in \{-1,0,\dots,|\word|\}$),
 $\ObservableConf{i}$ (resp.~$\UnobservableConf{i}$) is the set of 5-tuples $(\loc,\clockval,\tpval,\varval,\dpval)$ such that
 there is a timed data word $\word'$ over $\ActionsWithU$ satisfying the following:
 \begin{ienumeration}
	\item $(\locinit, \varvalo, \ClocksZero) \xrightarrow{\word'} (\loc, \varval, \clockval)$ in $\valuate{\A}{\dtpval{\tpval}{\dpval}}$,
	\item $\project{\word'}{\Actions} = \word(1,i)$,
	\item The last action $a'_{|\word'|}$ of $\word'$ is observable (resp.~unobservable and its timestamp is less than $\tau_{i+1}$).
 \end{ienumeration}
\end{definition}

\begin{algorithm}[tbp]
 \caption{Outline of our algorithm for symbolic monitoring}
 \label{algorithm:outline}
 %%%%% I can make this pseudo-code smaller but I did not because anyway it will be one-page
% \scalebox{1.0}{ 
% \parbox{1.0\textwidth}{
 \small
 \DontPrintSemicolon
 \newcommand{\myCommentFont}[1]{\texttt{\footnotesize{#1}}}
 \SetCommentSty{myCommentFont}
 \KwIn{A PTDA~$\A = (\ActionsWithU, \Loc, \locinit, \LocFinal, \Clock, \TParam, \Variables, \LVariables, \varvalo, \VParam, \Edges)$ over a \datatype{}~$(\Domain, \DomainGuards , \DomainUpdates)$ and actions $\ActionsWithU$,
 and a timed data word~$\word= (\action_1, \tau_1, \lvarval_1), (\action_2, \tau_2, \lvarval_2), \dots,(\action_n, \tau_n, \lvarval_n)$ over $\Actions$}
 \KwOut{$\bigcup_{i \in \{1,2,\dots,n+1\}} \Resulti{i}$ is the validity domain $\ValidityDomain{\word}{\A}$}
 $\UnobservableConf{-1} \gets \emptyset$;
 $\ObservableConf{0} \gets \{(\locinit,\ClocksZero, \tpval ,\varvalo, \dpval) \mid \tpval \in \PVal, \dpval \in \Domain^\VParam\}$\;
 \For{$i \gets 1$ \KwTo $n$}{
 \KwCompute $(\UnobservableConf{i-1}, \ObservableConf{i})$ \KwFrom $(\UnobservableConf{i-2}, \ObservableConf{i-1})$\;\nllabel{algo:compute}
 $\Resulti{i} \gets \{(\tpval,\dpval) \mid \exists (\loc,\clockval,\tpval,\varval,\dpval) \in \UnobservableConf{i-1} \cup \ObservableConf{i}.\, \loc\in\LocFinal \}$\;\nllabel{algo:accepting}
 }
 \KwCompute $\UnobservableConf{n}$ \KwFrom $(\UnobservableConf{n-1}, \ObservableConf{n})$\;\label{alg_line:outline:append_result}\nllabel{algo:compute-final}
 $\Resulti{n+1} \gets \{(\tpval,\dpval)  \mid \exists (\loc,\clockval,\tpval,\varval,\dpval) \in \UnobservableConf{n}.\, \loc\in\LocFinal \}$\nllabel{algo:accepting-final}
% }}
\end{algorithm}

\Cref{algorithm:outline} shows an outline of our algorithm for symbolic monitoring (see \LongVersion{\cref{appendix:algorithm_detail}}\ShortVersion{{\color{red} TODO: cite arXiv version}} for the full version).
%\ea{I would move this previous sentence after describing the notation, just before termination}
% EXPLANATION OF THE ALGORITHM
Our algorithm incrementally computes $\UnobservableConf{i-1}$ and $\ObservableConf{i}$ (\cref{algo:compute}).
After reading $(\action_i,\tau_i,\lvarval_i)$, our algorithm stores the partial results $(\tpval, \dpval) \in \ValidityDomain{\word}{\A}$ witnessed from the accepting configurations in
$\UnobservableConf{i-1}$ and $\ObservableConf{i}$  (\cref{algo:accepting}).
(We also need to try to take potential unobservable transitions and store the results from the accepting configurations \emph{after} the last element of the timed data word (\cref{algo:compute-final,algo:accepting-final}).)

% SYMBOLIC
%
Since $\PVal \times \Domain^{\VParam}$ is an infinite set, we cannot try each $(\tpval, \dpval) \in \PVal \times \Domain^{\VParam}$ and we use a symbolic representation for parameter valuations.
Similarly to the reachability synthesis of parametric timed automata~\cite{JLR15},
a set of clock and timing parameter valuations can be represented by a convex polyhedron.
For variable valuations and data parameter valuations, we need an appropriate representation depending on the \datatype{}~$(\Domain, \DomainGuards , \DomainUpdates)$.
Moreover, for the termination of \cref{algorithm:outline}, some operations on the symbolic representation are required.

\begin{theorem}[termination]
 For any PTDA~$\A$ over a \datatype{}~$(\Domain, \DomainGuards , \DomainUpdates)$ and actions $\ActionsWithU$, and for any timed data word~$\word$ over $\Actions$,
 \cref{algorithm:outline} terminates if the following operations on the symbolic representation $\vardpvals$ of a set of variable and data parameter valuations terminate.
 \begin{enumerate}
  \item restriction and update $\{ (\DPValuate{\PDU}{\dpval}{\varval}{\lvarval}, \dpval) \mid \exists (\varval,\dpval)\in \vardpvals.\, (\varval, \lvarval) \models \valuate{\dguard}{\dpval} \}$, 
        where $\lvarval$ is a local variable valuation, $\PDU$ is a parametric data update function, and $\dguard$ is a data guard;
  \item emptiness checking of $\vardpvals$;
  \item projection $\project{\vardpvals}{\VParam}$ of $\vardpvals$ to the data parameters $\VParam$.\qed
 \end{enumerate}
\end{theorem}

\begin{example}
 For the \datatype{} for rationals in \cref{example:data-types}, variable and data parameter valuations $\vardpvals$ can be represented by convex polyhedra and the above operations terminate.
 For the \datatype{} for strings $\grands$ in \cref{example:data-types}, variable and data parameter valuations $\vardpvals$ can be represented by $\grands^{|\Variables|} \times (\grands \cup \finitePowerset{\grands})^{|\VParam|}$ and the above operations terminate, where 
$\finitePowerset{\grands}$ is the set of finite sets of $\grands$.
\end{example}

\subsection{Encoding parametric timed pattern matching}\label{ss:PTPM}
%%%%%%%%%%%%%%%%%%%%%%%%%%%%%%%%%%%%%%%%%%%%%%%%%%%%%%%%%%%%

The symbolic monitoring problem is a generalization of the parametric timed pattern matching problem of~\cite{AHW18}.
Recall that parametric timed pattern matching aims at synthesizing timing parameter valuations and \emph{start and end times in the log} for which a log segment satisfies or violates a specification.
In our approach, by adding a clock measuring the absolute time, and two timing parameters encoding respectively the start and end date of the segment, one can easily infer the log segments for which the property is satisfied.
\LongVersion{%
	We note that even with \cref{assumption:no_unobservable_loop},  symbolic monitoring is still a generalization of parametric timed pattern matching.
}

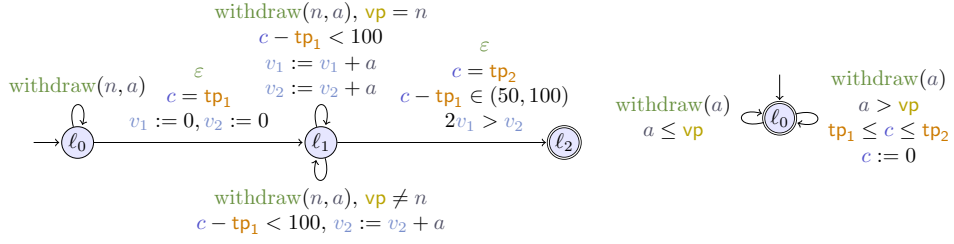
\begin{figure*}[t]
	\begin{subfigure}[h]{0.65\textwidth}
	\centering
		\scalebox{0.8}{
		\begin{tikzpicture}[shorten >=1pt,node distance=4.0cm,on grid,auto]
		%% states
		\node[location,initial] (s_0) {$\loc_0$};
		\node[location] (s_1) [right of=s_0,node distance=4.0cm] {$\loc_1$};
		\node[location,accepting] (s_2) [right of=s_1] {$\loc_2$};

		%% edges
		\path[->] 
		(s_0) edge [loop above] node {$\styleact{\mathrm{withdraw}}(\stylelvar{n},\stylelvar{a})$} (s_0)
		(s_0) edge [above] node {\begin{tabular}{c}
                                          $\styleact{\varepsilon}$\\
                                          $\styleclock{\clock} = \styleparam{\tparam_1}$\\
                                          $\stylevar{\dvar_1} := 0, \stylevar{\dvar_2} := 0$
                                         \end{tabular}} (s_1)
		(s_1) edge [loop above] node[align=center] {
                                               $\styleact{\mathrm{withdraw}}(\stylelvar{n},\stylelvar{a})$,
                                               $\styledparam{\vparam} = \stylelvar{n}$\\
                                               $\styleclock{\clock} - \styleparam{\styleparam{\tparam_1}} < 100$\\
                                               $\stylevar{\dvar_1} := \stylevar{\dvar_1} + \stylelvar{a}$\\
                                               $\stylevar{\dvar_2} := \stylevar{\dvar_2} + \stylelvar{a}$
                                              } (s_1)
		(s_1) edge [loop below] node[align=center] {
                                               $\styleact{\mathrm{withdraw}}(\stylelvar{n},\stylelvar{a})$,
                                               $\styledparam{\vparam} \neq \stylelvar{n}$\\
                                               $\styleclock{\clock} - \styleparam{\styleparam{\tparam_1}} < 100$,
                                               $\stylevar{\dvar_2} := \stylevar{\dvar_2} + \stylelvar{a}$
                                              } (s_1)
		(s_1) edge [above] node[pos=0.7] {\begin{tabular}{c}
                                          $\styleact{\varepsilon}$\\
                                          $\styleclock{\clock} = \styleparam{\styleparam{\tparam_2}}$\\
                                          $\styleclock{\clock} - \styleparam{\styleparam{\tparam_1}} \in (50,100)$\\
                                          $2\stylevar{\dvar_1} > \stylevar{\dvar_2}$
                                         \end{tabular}} (s_2);
		\end{tikzpicture}}
%	\caption{PTDA in \textsc{Dominant}}
%	\label{figure:patterns:dominant_withdraw}
	\end{subfigure}	
	%------------------------------------------------------------
	\begin{subfigure}[h]{0.33\textwidth}
	\centering
		\scalebox{0.8}{
		\begin{tikzpicture}[shorten >=1pt,node distance=2.5cm,on grid,auto]
		%% states
		\node[location,initial above,accepting] (s_0) {$\loc_0$};

		%% edges
		\path[->] 
		(s_0) edge [loop left] node {\begin{tabular}{c}
                                               $\styleact{\mathrm{withdraw}}(\stylelvar{a})$\\
                                               $\stylelvar{a} \leq \styledparam{\vparam}$\\
                                             \end{tabular}} (s_0)
		(s_0) edge [loop right] node {\begin{tabular}{c}
                                               $\styleact{\mathrm{withdraw}}(\stylelvar{a})$\\
                                               $\stylelvar{a} > \styledparam{\vparam}$\\
                                               $\styleparam{\tparam_1} \leq \styleclock{\clock} \leq \styleparam{\tparam_2}$\\
                                               $\styleclock{\clock} := 0$
                                              \end{tabular}} (s_0);
		\end{tikzpicture}}
%	\caption{PTDA in \textsc{Periodic}}
%	\label{figure:patterns:periodic_withdraw}
	\end{subfigure}

	\caption{PTDAs in \textsc{Dominant} (left) and \textsc{Periodic} (right)}
	\label{figure:patterns}
\end{figure*}

Consider the \textsc{Dominant} PTDA (left of \cref{figure:patterns}).
It is inspired by a monitoring of withdrawals from bank accounts of various users~\cite{BKZ17}.
This PTDA monitors situations when a user withdraws more than half of the total withdrawals within a time window of $(50, 100)$.
The actions are $\Actions = \{\styleact{withdraw}\}$ and $\ActDomain(\styleact{withdraw}) = \{n,a\}$, 
	where $n$ has a string value and $a$ has an integer value.
The string $n$ represents a user name and the integer $a$ represents the amount of the withdrawal by the user~$n$.
Observe that clock~$\styleclock{\clock}$ is never reset, and therefore measures absolute time.
The automaton can non-deterministically remain in~$\loc_0$, or start to measure a log by taking the $\uAction$-transition to~$\loc_1$ checking $\styleclock{\clock} = \styleparam{\tparam_1}$, and therefore ``remembering'' the start time using timing parameter~$\styleparam{\tparam_1}$.
Then, whenever a user~$\styledparam{vp}$ has withdrawn more than half of the accumulated withdrawals (data guard $2 \stylevar{\dvar_1} > \stylevar{\dvar_2}$) in a $(50, 100)$ time window (timed guard $\styleclock{\clock} - \styleparam{\styleparam{\tparam_1}} \in (50,100)$), the automaton takes a $\uAction$-transition to the accepting location, checking $\styleclock{\clock} = \styleparam{\tparam_2}$, and therefore remembering the end time using timing parameter~$\styleparam{\tparam_2}$.
%\ea{OK?}
	
%%%%%%%%%%%%%%%%%%%%%%%%%%%%%%%%%%%%%%%%%%%%%%%%%%%%%%%%%%%%
%%%%%%%%%%%%%%%%%%%%%%%%%%%%%%%%%%%%%%%%%%%%%%%%%%%%%%%%%%%%

%%%%%%%%%%%%%%%%%%%%%%%%%%%%%%%%%%%%%%%%%%%%%%%%%%%%%%%%%%%%
%%%%%%%%%%%%%%%%%%%%%%%%%%%%%%%%%%%%%%%%%%%%%%%%%%%%%%%%%%%%
\section{Experiments}\label{section:experiments}
%%%%%%%%%%%%%%%%%%%%%%%%%%%%%%%%%%%%%%%%%%%%%%%%%%%%%%%%%%%%
%%%%%%%%%%%%%%%%%%%%%%%%%%%%%%%%%%%%%%%%%%%%%%%%%%%%%%%%%%%%

%\ea{Masaki, can you add a (anonymous!) Web page or Web site where all data can be found? or is the code too not-anonymous?}
%\mw{I published \masakiTool{} at GitHub and added the link. I will also add the data to the repository later.}

We implemented our symbolic monitoring algorithm in a tool \masakiTool{} in C++\LongVersion{ (compiled using GCC 7.3.0)}, where the domain for data is the strings and the integers.\LongVersion{\footnote{%
	The use of integers is not an essential limitation. We may scale any rational number to an integer.
}}
\LongVersion{%
	For the strings, we used the \datatype{} in \cref{example:data-types} and 
	for integers, we used the \datatype{} for the rationals in \cref{example:data-types}, where any occurrences of $\grandq$ are replaced by $\grandz$.
}%
% \ea{I don't really understand; we defined rationals and strings. But you implemented integers and strings, right? So it's not really \cref{example:data-types}, is it?}
% \mw{You are right. I fixed it.}
Our tool \masakiTool{} is distributed at \url{https://github.com/MasWag/symon}.
We use \LongVersion{the Parma Polyhedra Library (}PPL\LongVersion{)}~\cite{BHZ08} for the symbolic representation of the valuations. %(\eg{} $\clocktpvals$ and $\vardpvals$ in \cref{algorithm}).
We note that we employ an optimization to merge adjacent polyhedra in the configurations if possible.\LongVersion{\footnote{%
%	\ea{when do you do this? When computing a new state? Do you do it recursively?}\mw{I do it after consuming each entry $(\action_i,\tau_i,\lvarval_i)$ of the timed word $\word$ \ie{} in \cref{alg_line:set_next_curr_conf} of \cref{algorithm}. I use \texttt{Pointset\_Powerset::pairwise\_reduce}.}
	After consuming each entry $(\action_i,\tau_i,\lvarval_i)$ of the timed word $\word$ (\ie{} in \cref{alg_line:outline:append_result} of \cref{algorithm:outline}), we use PPL's \texttt{Pointset\_Powerset::pairwise\_reduce} function.%\ea{added this as it's important for long-term understanding of our implementation (when calling this kind of function is important!)}
}}
We evaluated our monitor algorithm against three original benchmarks: \LongVersion{the PTDA in }\textsc{Copy}\LongVersion{ is} in \cref{figure:example:copy:automaton}; and \LongVersion{the PTDAs in }\textsc{Dominant} and \textsc{Periodic} \LongVersion{are shown }in \cref{figure:patterns}.
\LongVersion{

}%
We conducted \LongVersion{the }experiments on an Amazon EC2 c4.large instance (2.9\,GHz Intel Xeon E5-2666 v3, 2 vCPUs, and 3.75\,GiB RAM) that runs Ubuntu 18.04 LTS (64\,bit).
%Experiment data can be found on \todo{NOT YET!!!}.

% \todo{Merge two plots if we do not compare with \imitator}

\subsection{Benchmark 1: \textsc{Copy}}

Our first benchmark \textsc{Copy} is a monitoring of variable updates much like the scenario in \cite{BDSV14}. %\mw{I could not find something similar in \cite{BDSV14}. Is there any reference?}\ea{I think that's \cite[p.6]{BDSV14}}\mw{I see. Thanks.}
The actions are $\Actions = \{\styleact{update}\}$ and $\ActDomain(\styleact{update}) = \{n, v\}$,
where $n$ has a string value representing the name of the updated variables and $v$ has an integer value representing the updated value.
\LongVersion{%
	We generated random timed data words of various sizes.
}%
Our set \LongVersion{$\words$ }consists of 10 timed data words of length 4,000 to 40,000.

The PTDA in \textsc{Copy} is shown in \cref{figure:example:copy:automaton}, where we give an additional constraint $3 < \tparam< 10$ on $\tparam$.
The property encoded in \cref{figure:example:copy:automaton} is ``for any variable~$\styledparam{px}$, whenever an update of that variable occurs, then within $\tparam$ time units, the value of \texttt{b} must be equal to that update''.

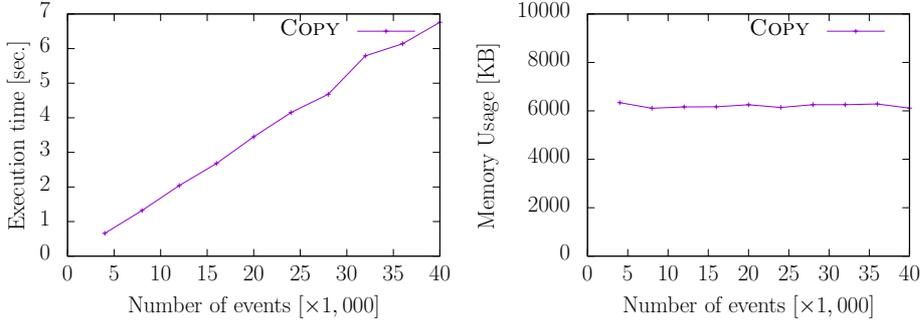
\begin{figure}[tbp]
 \begin{minipage}{0.5\linewidth}
  \scalebox{0.50}{\begin{tikzpicture}[gnuplot]
%% generated with GNUPLOT 5.2p6 (Lua 5.3; terminal rev. Nov 2018, script rev. 107)
%% Fri Feb  8 15:01:12 2019
\tikzset{every node/.append style={font={\fontsize{18.0pt}{21.6pt}\selectfont}}}
\path (0.000,0.000) rectangle (12.500,8.750);
\gpcolor{color=gp lt color border}
\gpsetlinetype{gp lt border}
\gpsetdashtype{gp dt solid}
\gpsetlinewidth{1.00}
\draw[gp path] (1.712,1.772)--(1.892,1.772);
\draw[gp path] (11.506,1.772)--(11.326,1.772);
\node[gp node right] at (1.381,1.772) {$0$};
\draw[gp path] (1.712,2.690)--(1.892,2.690);
\draw[gp path] (11.506,2.690)--(11.326,2.690);
\node[gp node right] at (1.381,2.690) {$1$};
\draw[gp path] (1.712,3.607)--(1.892,3.607);
\draw[gp path] (11.506,3.607)--(11.326,3.607);
\node[gp node right] at (1.381,3.607) {$2$};
\draw[gp path] (1.712,4.525)--(1.892,4.525);
\draw[gp path] (11.506,4.525)--(11.326,4.525);
\node[gp node right] at (1.381,4.525) {$3$};
\draw[gp path] (1.712,5.442)--(1.892,5.442);
\draw[gp path] (11.506,5.442)--(11.326,5.442);
\node[gp node right] at (1.381,5.442) {$4$};
\draw[gp path] (1.712,6.360)--(1.892,6.360);
\draw[gp path] (11.506,6.360)--(11.326,6.360);
\node[gp node right] at (1.381,6.360) {$5$};
\draw[gp path] (1.712,7.277)--(1.892,7.277);
\draw[gp path] (11.506,7.277)--(11.326,7.277);
\node[gp node right] at (1.381,7.277) {$6$};
\draw[gp path] (1.712,8.195)--(1.892,8.195);
\draw[gp path] (11.506,8.195)--(11.326,8.195);
\node[gp node right] at (1.381,8.195) {$7$};
\draw[gp path] (1.712,1.772)--(1.712,1.952);
\draw[gp path] (1.712,8.195)--(1.712,8.015);
\node[gp node center] at (1.712,1.218) {$0$};
\draw[gp path] (2.936,1.772)--(2.936,1.952);
\draw[gp path] (2.936,8.195)--(2.936,8.015);
\node[gp node center] at (2.936,1.218) {$5$};
\draw[gp path] (4.161,1.772)--(4.161,1.952);
\draw[gp path] (4.161,8.195)--(4.161,8.015);
\node[gp node center] at (4.161,1.218) {$10$};
\draw[gp path] (5.385,1.772)--(5.385,1.952);
\draw[gp path] (5.385,8.195)--(5.385,8.015);
\node[gp node center] at (5.385,1.218) {$15$};
\draw[gp path] (6.609,1.772)--(6.609,1.952);
\draw[gp path] (6.609,8.195)--(6.609,8.015);
\node[gp node center] at (6.609,1.218) {$20$};
\draw[gp path] (7.833,1.772)--(7.833,1.952);
\draw[gp path] (7.833,8.195)--(7.833,8.015);
\node[gp node center] at (7.833,1.218) {$25$};
\draw[gp path] (9.058,1.772)--(9.058,1.952);
\draw[gp path] (9.058,8.195)--(9.058,8.015);
\node[gp node center] at (9.058,1.218) {$30$};
\draw[gp path] (10.282,1.772)--(10.282,1.952);
\draw[gp path] (10.282,8.195)--(10.282,8.015);
\node[gp node center] at (10.282,1.218) {$35$};
\draw[gp path] (11.506,1.772)--(11.506,1.952);
\draw[gp path] (11.506,8.195)--(11.506,8.015);
\node[gp node center] at (11.506,1.218) {$40$};
\draw[gp path] (1.712,8.195)--(1.712,1.772)--(11.506,1.772)--(11.506,8.195)--cycle;
\node[gp node center,rotate=-270] at (0.496,4.983) {Execution time [sec.]};
\node[gp node center] at (6.609,0.387) {Number of events [$\times 1,000$]};
\node[gp node right] at (9.009,7.738) {\textsc{Copy}};
\gpcolor{rgb color={0.580,0.000,0.827}}
\draw[gp path] (9.340,7.738)--(10.844,7.738);
\draw[gp path] (2.691,2.378)--(3.671,2.983)--(4.650,3.644)--(5.630,4.231)--(6.609,4.938)%
  --(7.588,5.580)--(8.568,6.066)--(9.547,7.085)--(10.527,7.406)--(11.506,7.975);
\gpsetpointsize{4.00}
\gppoint{gp mark 1}{(2.691,2.378)}
\gppoint{gp mark 1}{(3.671,2.983)}
\gppoint{gp mark 1}{(4.650,3.644)}
\gppoint{gp mark 1}{(5.630,4.231)}
\gppoint{gp mark 1}{(6.609,4.938)}
\gppoint{gp mark 1}{(7.588,5.580)}
\gppoint{gp mark 1}{(8.568,6.066)}
\gppoint{gp mark 1}{(9.547,7.085)}
\gppoint{gp mark 1}{(10.527,7.406)}
\gppoint{gp mark 1}{(11.506,7.975)}
\gppoint{gp mark 1}{(10.092,7.738)}
\gpcolor{color=gp lt color border}
\draw[gp path] (1.712,8.195)--(1.712,1.772)--(11.506,1.772)--(11.506,8.195)--cycle;
%% coordinates of the plot area
\gpdefrectangularnode{gp plot 1}{\pgfpoint{1.712cm}{1.772cm}}{\pgfpoint{11.506cm}{8.195cm}}
\end{tikzpicture}
%% gnuplot variables
}
 \end{minipage}
 \begin{minipage}{0.5\linewidth}
  \scalebox{0.50}{\begin{tikzpicture}[gnuplot]
%% generated with GNUPLOT 5.2p6 (Lua 5.3; terminal rev. Nov 2018, script rev. 107)
%% Fri Feb  8 15:01:12 2019
\tikzset{every node/.append style={font={\fontsize{18.0pt}{21.6pt}\selectfont}}}
\path (0.000,0.000) rectangle (12.500,8.750);
\gpcolor{color=gp lt color border}
\gpsetlinetype{gp lt border}
\gpsetdashtype{gp dt solid}
\gpsetlinewidth{1.00}
\draw[gp path] (3.036,1.772)--(3.216,1.772);
\draw[gp path] (11.506,1.772)--(11.326,1.772);
\node[gp node right] at (2.705,1.772) {$0$};
\draw[gp path] (3.036,3.057)--(3.216,3.057);
\draw[gp path] (11.506,3.057)--(11.326,3.057);
\node[gp node right] at (2.705,3.057) {$2000$};
\draw[gp path] (3.036,4.341)--(3.216,4.341);
\draw[gp path] (11.506,4.341)--(11.326,4.341);
\node[gp node right] at (2.705,4.341) {$4000$};
\draw[gp path] (3.036,5.626)--(3.216,5.626);
\draw[gp path] (11.506,5.626)--(11.326,5.626);
\node[gp node right] at (2.705,5.626) {$6000$};
\draw[gp path] (3.036,6.910)--(3.216,6.910);
\draw[gp path] (11.506,6.910)--(11.326,6.910);
\node[gp node right] at (2.705,6.910) {$8000$};
\draw[gp path] (3.036,8.195)--(3.216,8.195);
\draw[gp path] (11.506,8.195)--(11.326,8.195);
\node[gp node right] at (2.705,8.195) {$10000$};
\draw[gp path] (3.036,1.772)--(3.036,1.952);
\draw[gp path] (3.036,8.195)--(3.036,8.015);
\node[gp node center] at (3.036,1.218) {$0$};
\draw[gp path] (4.095,1.772)--(4.095,1.952);
\draw[gp path] (4.095,8.195)--(4.095,8.015);
\node[gp node center] at (4.095,1.218) {$5$};
\draw[gp path] (5.154,1.772)--(5.154,1.952);
\draw[gp path] (5.154,8.195)--(5.154,8.015);
\node[gp node center] at (5.154,1.218) {$10$};
\draw[gp path] (6.212,1.772)--(6.212,1.952);
\draw[gp path] (6.212,8.195)--(6.212,8.015);
\node[gp node center] at (6.212,1.218) {$15$};
\draw[gp path] (7.271,1.772)--(7.271,1.952);
\draw[gp path] (7.271,8.195)--(7.271,8.015);
\node[gp node center] at (7.271,1.218) {$20$};
\draw[gp path] (8.330,1.772)--(8.330,1.952);
\draw[gp path] (8.330,8.195)--(8.330,8.015);
\node[gp node center] at (8.330,1.218) {$25$};
\draw[gp path] (9.389,1.772)--(9.389,1.952);
\draw[gp path] (9.389,8.195)--(9.389,8.015);
\node[gp node center] at (9.389,1.218) {$30$};
\draw[gp path] (10.447,1.772)--(10.447,1.952);
\draw[gp path] (10.447,8.195)--(10.447,8.015);
\node[gp node center] at (10.447,1.218) {$35$};
\draw[gp path] (11.506,1.772)--(11.506,1.952);
\draw[gp path] (11.506,8.195)--(11.506,8.015);
\node[gp node center] at (11.506,1.218) {$40$};
\draw[gp path] (3.036,8.195)--(3.036,1.772)--(11.506,1.772)--(11.506,8.195)--cycle;
\node[gp node center,rotate=-270] at (0.496,4.983) {Memory Usage [KB]};
\node[gp node center] at (7.271,0.387) {Number of events [$\times 1,000$]};
\node[gp node right] at (9.009,7.738) {\textsc{Copy}};
\gpcolor{rgb color={0.580,0.000,0.827}}
\draw[gp path] (9.340,7.738)--(10.844,7.738);
\draw[gp path] (3.883,5.844)--(4.730,5.695)--(5.577,5.731)--(6.424,5.734)--(7.271,5.788)%
  --(8.118,5.716)--(8.965,5.790)--(9.812,5.790)--(10.659,5.808)--(11.506,5.698);
\gpsetpointsize{4.00}
\gppoint{gp mark 1}{(3.883,5.844)}
\gppoint{gp mark 1}{(4.730,5.695)}
\gppoint{gp mark 1}{(5.577,5.731)}
\gppoint{gp mark 1}{(6.424,5.734)}
\gppoint{gp mark 1}{(7.271,5.788)}
\gppoint{gp mark 1}{(8.118,5.716)}
\gppoint{gp mark 1}{(8.965,5.790)}
\gppoint{gp mark 1}{(9.812,5.790)}
\gppoint{gp mark 1}{(10.659,5.808)}
\gppoint{gp mark 1}{(11.506,5.698)}
\gppoint{gp mark 1}{(10.092,7.738)}
\gpcolor{color=gp lt color border}
\draw[gp path] (3.036,8.195)--(3.036,1.772)--(11.506,1.772)--(11.506,8.195)--cycle;
%% coordinates of the plot area
\gpdefrectangularnode{gp plot 1}{\pgfpoint{3.036cm}{1.772cm}}{\pgfpoint{11.506cm}{8.195cm}}
\end{tikzpicture}
%% gnuplot variables
}
 \end{minipage}
 \caption{Execution time (left) and memory usage (right) of \textsc{Copy}}
 \label{figure:result:copy}
\end{figure}

The experiment result is in \cref{figure:result:copy}.
We observe that the execution time is linear to the number of the events and the memory usage is more or less constant with respect to the number of events.

%\mw{It is quite fast. This is because the constraint $3 < \tparam< 10$ is strict and the number of configurations is very small. If I change it to $3 < \tparam< 100$, the execution time gets about 5 times longer.}
%\ea{good! Perhaps it would be good to output a (small) part of the answer. Because, in the current manuscript, there is no way for the reader to understand what we mean with ``symbolic''}

\subsection{Benchmark 2: \textsc{Dominant}}

\begin{figure}[tbp]
 \begin{minipage}{0.5\linewidth}
  \scalebox{0.50}{\begin{tikzpicture}[gnuplot]
%% generated with GNUPLOT 5.2p6 (Lua 5.3; terminal rev. Nov 2018, script rev. 107)
%% Fri Feb  8 14:57:29 2019
\tikzset{every node/.append style={font={\fontsize{18.0pt}{21.6pt}\selectfont}}}
\path (0.000,0.000) rectangle (12.500,8.750);
\gpcolor{color=gp lt color border}
\gpsetlinetype{gp lt border}
\gpsetdashtype{gp dt solid}
\gpsetlinewidth{1.00}
\draw[gp path] (2.374,1.772)--(2.554,1.772);
\draw[gp path] (11.506,1.772)--(11.326,1.772);
\node[gp node right] at (2.043,1.772) {$0$};
\draw[gp path] (2.374,2.575)--(2.554,2.575);
\draw[gp path] (11.506,2.575)--(11.326,2.575);
\node[gp node right] at (2.043,2.575) {$20$};
\draw[gp path] (2.374,3.378)--(2.554,3.378);
\draw[gp path] (11.506,3.378)--(11.326,3.378);
\node[gp node right] at (2.043,3.378) {$40$};
\draw[gp path] (2.374,4.181)--(2.554,4.181);
\draw[gp path] (11.506,4.181)--(11.326,4.181);
\node[gp node right] at (2.043,4.181) {$60$};
\draw[gp path] (2.374,4.984)--(2.554,4.984);
\draw[gp path] (11.506,4.984)--(11.326,4.984);
\node[gp node right] at (2.043,4.984) {$80$};
\draw[gp path] (2.374,5.786)--(2.554,5.786);
\draw[gp path] (11.506,5.786)--(11.326,5.786);
\node[gp node right] at (2.043,5.786) {$100$};
\draw[gp path] (2.374,6.589)--(2.554,6.589);
\draw[gp path] (11.506,6.589)--(11.326,6.589);
\node[gp node right] at (2.043,6.589) {$120$};
\draw[gp path] (2.374,7.392)--(2.554,7.392);
\draw[gp path] (11.506,7.392)--(11.326,7.392);
\node[gp node right] at (2.043,7.392) {$140$};
\draw[gp path] (2.374,8.195)--(2.554,8.195);
\draw[gp path] (11.506,8.195)--(11.326,8.195);
\node[gp node right] at (2.043,8.195) {$160$};
\draw[gp path] (2.374,1.772)--(2.374,1.952);
\draw[gp path] (2.374,8.195)--(2.374,8.015);
\node[gp node center] at (2.374,1.218) {$2$};
\draw[gp path] (3.389,1.772)--(3.389,1.952);
\draw[gp path] (3.389,8.195)--(3.389,8.015);
\node[gp node center] at (3.389,1.218) {$4$};
\draw[gp path] (4.403,1.772)--(4.403,1.952);
\draw[gp path] (4.403,8.195)--(4.403,8.015);
\node[gp node center] at (4.403,1.218) {$6$};
\draw[gp path] (5.418,1.772)--(5.418,1.952);
\draw[gp path] (5.418,8.195)--(5.418,8.015);
\node[gp node center] at (5.418,1.218) {$8$};
\draw[gp path] (6.433,1.772)--(6.433,1.952);
\draw[gp path] (6.433,8.195)--(6.433,8.015);
\node[gp node center] at (6.433,1.218) {$10$};
\draw[gp path] (7.447,1.772)--(7.447,1.952);
\draw[gp path] (7.447,8.195)--(7.447,8.015);
\node[gp node center] at (7.447,1.218) {$12$};
\draw[gp path] (8.462,1.772)--(8.462,1.952);
\draw[gp path] (8.462,8.195)--(8.462,8.015);
\node[gp node center] at (8.462,1.218) {$14$};
\draw[gp path] (9.477,1.772)--(9.477,1.952);
\draw[gp path] (9.477,8.195)--(9.477,8.015);
\node[gp node center] at (9.477,1.218) {$16$};
\draw[gp path] (10.491,1.772)--(10.491,1.952);
\draw[gp path] (10.491,8.195)--(10.491,8.015);
\node[gp node center] at (10.491,1.218) {$18$};
\draw[gp path] (11.506,1.772)--(11.506,1.952);
\draw[gp path] (11.506,8.195)--(11.506,8.015);
\node[gp node center] at (11.506,1.218) {$20$};
\draw[gp path] (2.374,8.195)--(2.374,1.772)--(11.506,1.772)--(11.506,8.195)--cycle;
\node[gp node center,rotate=-270] at (0.496,4.983) {Execution time [sec.]};
\node[gp node center] at (6.940,0.387) {Number of events [$\times 1,000$]};
\node[gp node right] at (9.009,7.738) {\textsc{Dominant}};
\gpcolor{rgb color={0.580,0.000,0.827}}
\draw[gp path] (9.340,7.738)--(10.844,7.738);
\draw[gp path] (2.374,2.360)--(3.389,2.945)--(4.403,3.555)--(5.418,4.252)--(6.433,4.821)%
  --(7.447,5.290)--(8.462,6.065)--(9.477,6.658)--(10.491,7.089)--(11.506,7.722);
\gpsetpointsize{4.00}
\gppoint{gp mark 1}{(2.374,2.360)}
\gppoint{gp mark 1}{(3.389,2.945)}
\gppoint{gp mark 1}{(4.403,3.555)}
\gppoint{gp mark 1}{(5.418,4.252)}
\gppoint{gp mark 1}{(6.433,4.821)}
\gppoint{gp mark 1}{(7.447,5.290)}
\gppoint{gp mark 1}{(8.462,6.065)}
\gppoint{gp mark 1}{(9.477,6.658)}
\gppoint{gp mark 1}{(10.491,7.089)}
\gppoint{gp mark 1}{(11.506,7.722)}
\gppoint{gp mark 1}{(10.092,7.738)}
\gpcolor{color=gp lt color border}
\node[gp node right] at (9.009,7.184) {\textsc{Periodic}};
\gpcolor{rgb color={0.000,0.620,0.451}}
\draw[gp path] (9.340,7.184)--(10.844,7.184);
\draw[gp path] (2.374,2.039)--(3.389,2.371)--(4.403,2.447)--(5.418,2.890)--(6.433,3.243)%
  --(7.447,3.281)--(8.462,4.017)--(9.477,4.064)--(10.491,4.242)--(11.506,4.566);
\gppoint{gp mark 2}{(2.374,2.039)}
\gppoint{gp mark 2}{(3.389,2.371)}
\gppoint{gp mark 2}{(4.403,2.447)}
\gppoint{gp mark 2}{(5.418,2.890)}
\gppoint{gp mark 2}{(6.433,3.243)}
\gppoint{gp mark 2}{(7.447,3.281)}
\gppoint{gp mark 2}{(8.462,4.017)}
\gppoint{gp mark 2}{(9.477,4.064)}
\gppoint{gp mark 2}{(10.491,4.242)}
\gppoint{gp mark 2}{(11.506,4.566)}
\gppoint{gp mark 2}{(10.092,7.184)}
\gpcolor{color=gp lt color border}
\draw[gp path] (2.374,8.195)--(2.374,1.772)--(11.506,1.772)--(11.506,8.195)--cycle;
%% coordinates of the plot area
\gpdefrectangularnode{gp plot 1}{\pgfpoint{2.374cm}{1.772cm}}{\pgfpoint{11.506cm}{8.195cm}}
\end{tikzpicture}
%% gnuplot variables
}
 \end{minipage}
 \begin{minipage}{0.5\linewidth}
  \scalebox{0.50}{\begin{tikzpicture}[gnuplot]
%% generated with GNUPLOT 5.2p6 (Lua 5.3; terminal rev. Nov 2018, script rev. 107)
%% Fri Feb  8 14:57:29 2019
\tikzset{every node/.append style={font={\fontsize{18.0pt}{21.6pt}\selectfont}}}
\path (0.000,0.000) rectangle (12.500,8.750);
\gpcolor{color=gp lt color border}
\gpsetlinetype{gp lt border}
\gpsetdashtype{gp dt solid}
\gpsetlinewidth{1.00}
\draw[gp path] (3.036,1.772)--(3.216,1.772);
\draw[gp path] (11.506,1.772)--(11.326,1.772);
\node[gp node right] at (2.705,1.772) {$0$};
\draw[gp path] (3.036,3.057)--(3.216,3.057);
\draw[gp path] (11.506,3.057)--(11.326,3.057);
\node[gp node right] at (2.705,3.057) {$2000$};
\draw[gp path] (3.036,4.341)--(3.216,4.341);
\draw[gp path] (11.506,4.341)--(11.326,4.341);
\node[gp node right] at (2.705,4.341) {$4000$};
\draw[gp path] (3.036,5.626)--(3.216,5.626);
\draw[gp path] (11.506,5.626)--(11.326,5.626);
\node[gp node right] at (2.705,5.626) {$6000$};
\draw[gp path] (3.036,6.910)--(3.216,6.910);
\draw[gp path] (11.506,6.910)--(11.326,6.910);
\node[gp node right] at (2.705,6.910) {$8000$};
\draw[gp path] (3.036,8.195)--(3.216,8.195);
\draw[gp path] (11.506,8.195)--(11.326,8.195);
\node[gp node right] at (2.705,8.195) {$10000$};
\draw[gp path] (3.036,1.772)--(3.036,1.952);
\draw[gp path] (3.036,8.195)--(3.036,8.015);
\node[gp node center] at (3.036,1.218) {$2$};
\draw[gp path] (3.977,1.772)--(3.977,1.952);
\draw[gp path] (3.977,8.195)--(3.977,8.015);
\node[gp node center] at (3.977,1.218) {$4$};
\draw[gp path] (4.918,1.772)--(4.918,1.952);
\draw[gp path] (4.918,8.195)--(4.918,8.015);
\node[gp node center] at (4.918,1.218) {$6$};
\draw[gp path] (5.859,1.772)--(5.859,1.952);
\draw[gp path] (5.859,8.195)--(5.859,8.015);
\node[gp node center] at (5.859,1.218) {$8$};
\draw[gp path] (6.800,1.772)--(6.800,1.952);
\draw[gp path] (6.800,8.195)--(6.800,8.015);
\node[gp node center] at (6.800,1.218) {$10$};
\draw[gp path] (7.742,1.772)--(7.742,1.952);
\draw[gp path] (7.742,8.195)--(7.742,8.015);
\node[gp node center] at (7.742,1.218) {$12$};
\draw[gp path] (8.683,1.772)--(8.683,1.952);
\draw[gp path] (8.683,8.195)--(8.683,8.015);
\node[gp node center] at (8.683,1.218) {$14$};
\draw[gp path] (9.624,1.772)--(9.624,1.952);
\draw[gp path] (9.624,8.195)--(9.624,8.015);
\node[gp node center] at (9.624,1.218) {$16$};
\draw[gp path] (10.565,1.772)--(10.565,1.952);
\draw[gp path] (10.565,8.195)--(10.565,8.015);
\node[gp node center] at (10.565,1.218) {$18$};
\draw[gp path] (11.506,1.772)--(11.506,1.952);
\draw[gp path] (11.506,8.195)--(11.506,8.015);
\node[gp node center] at (11.506,1.218) {$20$};
\draw[gp path] (3.036,8.195)--(3.036,1.772)--(11.506,1.772)--(11.506,8.195)--cycle;
\node[gp node center,rotate=-270] at (0.496,4.983) {Memory Usage [KB]};
\node[gp node center] at (7.271,0.387) {Number of events [$\times 1,000$]};
\node[gp node right] at (9.009,7.738) {\textsc{Dominant}};
\gpcolor{rgb color={0.580,0.000,0.827}}
\draw[gp path] (9.340,7.738)--(10.844,7.738);
\draw[gp path] (3.036,6.222)--(3.977,6.245)--(4.918,6.245)--(5.859,6.240)--(6.800,6.227)%
  --(7.742,6.289)--(8.683,6.258)--(9.624,6.235)--(10.565,6.237)--(11.506,6.420);
\gpsetpointsize{4.00}
\gppoint{gp mark 1}{(3.036,6.222)}
\gppoint{gp mark 1}{(3.977,6.245)}
\gppoint{gp mark 1}{(4.918,6.245)}
\gppoint{gp mark 1}{(5.859,6.240)}
\gppoint{gp mark 1}{(6.800,6.227)}
\gppoint{gp mark 1}{(7.742,6.289)}
\gppoint{gp mark 1}{(8.683,6.258)}
\gppoint{gp mark 1}{(9.624,6.235)}
\gppoint{gp mark 1}{(10.565,6.237)}
\gppoint{gp mark 1}{(11.506,6.420)}
\gppoint{gp mark 1}{(10.092,7.738)}
\gpcolor{color=gp lt color border}
\node[gp node right] at (9.009,7.184) {\textsc{Periodic}};
\gpcolor{rgb color={0.000,0.620,0.451}}
\draw[gp path] (9.340,7.184)--(10.844,7.184);
\draw[gp path] (3.036,5.880)--(3.977,5.929)--(4.918,5.839)--(5.859,5.872)--(6.800,5.991)%
  --(7.742,5.988)--(8.683,5.865)--(9.624,5.942)--(10.565,5.908)--(11.506,5.872);
\gppoint{gp mark 2}{(3.036,5.880)}
\gppoint{gp mark 2}{(3.977,5.929)}
\gppoint{gp mark 2}{(4.918,5.839)}
\gppoint{gp mark 2}{(5.859,5.872)}
\gppoint{gp mark 2}{(6.800,5.991)}
\gppoint{gp mark 2}{(7.742,5.988)}
\gppoint{gp mark 2}{(8.683,5.865)}
\gppoint{gp mark 2}{(9.624,5.942)}
\gppoint{gp mark 2}{(10.565,5.908)}
\gppoint{gp mark 2}{(11.506,5.872)}
\gppoint{gp mark 2}{(10.092,7.184)}
\gpcolor{color=gp lt color border}
\draw[gp path] (3.036,8.195)--(3.036,1.772)--(11.506,1.772)--(11.506,8.195)--cycle;
%% coordinates of the plot area
\gpdefrectangularnode{gp plot 1}{\pgfpoint{3.036cm}{1.772cm}}{\pgfpoint{11.506cm}{8.195cm}}
\end{tikzpicture}
%% gnuplot variables
}
 \end{minipage}
 \caption{\small Execution time (left) and memory usage (right) of \textsc{Dominant} and \textsc{Periodic}}
 \label{figure:result:dominant-periodic}
\end{figure}
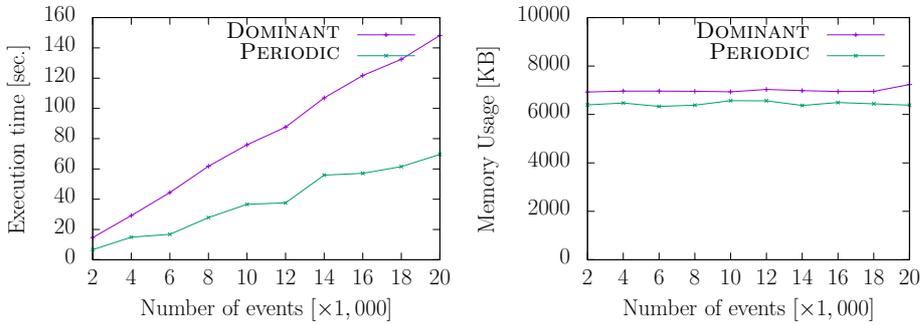

Our second benchmark is \textsc{Dominant} (\cref{figure:patterns} left).
\LongVersion{%
	We generated random timed data words of various sizes, where the number of users is 3 and the duration between each withdrawal follows the uniform distribution on $\{1,2,\dots,10\}$.
}%
Our set \LongVersion{$\words$ }consists of 10 timed data words of length 2,000 to 20,000.
\LongVersion{%
	Recall that this PTDA matches a situation when the amount of the withdrawal by the user $\vparam$ in a certain time window is more than the half of the withdrawals by all of the users in the same time window.
	The time window must be between 50 and 100.
	The parameters $\tparam_1$ and $\tparam_2$ show the beginning and the end of the time window respectively.

}%
The experiment result is in \cref{figure:result:dominant-periodic}.
We observe that the execution time is linear to the number of the events and the memory usage is more or less constant with respect to the number of events.
% 	\ea{can you use a scale from~0 for memory usage? Because the graphics shows much variation while it's in fact more or less constant}\mw{I did it.}

%\begin{figure}[tbp]
% \begin{minipage}{0.5\linewidth}
%  \scalebox{0.50}{  \input{./fig/dominant_withdraw_pattern_matching-time.tikz}}
% \end{minipage} 
% \begin{minipage}{0.5\linewidth}
%  \scalebox{0.50}{\input{./fig/dominant_withdraw_pattern_matching-ram.tikz}}
% \end{minipage}
% \caption{Execution time (left) and memory usage (right) of \textsc{Dominant}}
% \label{figure:result:dominant_withdraw}
%\end{figure}

\subsection{Benchmark 3: \textsc{Periodic}}

Our third benchmark \textsc{Periodic} is inspired by a parameter identification of periodic withdrawals from one bank account.
The actions are $\Actions = \{\styleact{withdraw}\}$ and $\ActDomain(\styleact{withdraw}) = \{a\}$,
where $a$ has an integer value representing the amount of the withdrawal.
We randomly generated a set \LongVersion{$\words$ }consisting of 10 timed data words of length 2,000 to 20,000.
Each timed data word consists of the following three kinds of periodic withdrawals:
%We added random numbers to both the timestamp and the amount of the withdrawal.

% \vspace{-1em}

\begin{description}
 \item[short period] One withdrawal occurs every $5\pm 1$ time units. The amount of the withdrawal is $50\pm 3$.
 \item[middle period] One withdrawal occurs every $50\pm 3$ time units. The amount of the withdrawal is $1000\pm 40$.
 \item[long period] One withdrawal occurs every $100\pm 5$ time units. The amount of the withdrawal is $5000\pm 20$.
\end{description}

\begin{wrapfigure}{r}{0pt}
 \scalebox{0.40}{\begin{tikzpicture}[gnuplot]
%% generated with GNUPLOT 5.2p6 (Lua 5.3; terminal rev. Nov 2018, script rev. 107)
%% Thu Apr 25 13:14:55 2019
\tikzset{every node/.append style={font={\fontsize{20.0pt}{24.0pt}\selectfont}}}
\path (0.000,0.000) rectangle (12.500,8.750);
\gpcolor{color=gp lt color axes}
\gpsetlinetype{gp lt axes}
\gpsetdashtype{gp dt axes}
\gpsetlinewidth{0.50}
\draw[gp path] (2.640,1.971)--(11.395,1.971);
\gpcolor{color=gp lt color border}
\gpsetlinetype{gp lt border}
\gpsetdashtype{gp dt solid}
\gpsetlinewidth{1.00}
\draw[gp path] (2.640,1.971)--(2.820,1.971);
\draw[gp path] (11.395,1.971)--(11.215,1.971);
\node[gp node right] at (2.272,1.971) {$0$};
\gpcolor{color=gp lt color axes}
\gpsetlinetype{gp lt axes}
\gpsetdashtype{gp dt axes}
\gpsetlinewidth{0.50}
\draw[gp path] (2.640,3.091)--(11.395,3.091);
\gpcolor{color=gp lt color border}
\gpsetlinetype{gp lt border}
\gpsetdashtype{gp dt solid}
\gpsetlinewidth{1.00}
\draw[gp path] (2.640,3.091)--(2.820,3.091);
\draw[gp path] (11.395,3.091)--(11.215,3.091);
\node[gp node right] at (2.272,3.091) {$20$};
\gpcolor{color=gp lt color axes}
\gpsetlinetype{gp lt axes}
\gpsetdashtype{gp dt axes}
\gpsetlinewidth{0.50}
\draw[gp path] (2.640,4.212)--(11.395,4.212);
\gpcolor{color=gp lt color border}
\gpsetlinetype{gp lt border}
\gpsetdashtype{gp dt solid}
\gpsetlinewidth{1.00}
\draw[gp path] (2.640,4.212)--(2.820,4.212);
\draw[gp path] (11.395,4.212)--(11.215,4.212);
\node[gp node right] at (2.272,4.212) {$40$};
\gpcolor{color=gp lt color axes}
\gpsetlinetype{gp lt axes}
\gpsetdashtype{gp dt axes}
\gpsetlinewidth{0.50}
\draw[gp path] (2.640,5.332)--(11.395,5.332);
\gpcolor{color=gp lt color border}
\gpsetlinetype{gp lt border}
\gpsetdashtype{gp dt solid}
\gpsetlinewidth{1.00}
\draw[gp path] (2.640,5.332)--(2.820,5.332);
\draw[gp path] (11.395,5.332)--(11.215,5.332);
\node[gp node right] at (2.272,5.332) {$60$};
\gpcolor{color=gp lt color axes}
\gpsetlinetype{gp lt axes}
\gpsetdashtype{gp dt axes}
\gpsetlinewidth{0.50}
\draw[gp path] (2.640,6.452)--(6.615,6.452);
\draw[gp path] (11.027,6.452)--(11.395,6.452);
\gpcolor{color=gp lt color border}
\gpsetlinetype{gp lt border}
\gpsetdashtype{gp dt solid}
\gpsetlinewidth{1.00}
\draw[gp path] (2.640,6.452)--(2.820,6.452);
\draw[gp path] (11.395,6.452)--(11.215,6.452);
\node[gp node right] at (2.272,6.452) {$80$};
\gpcolor{color=gp lt color axes}
\gpsetlinetype{gp lt axes}
\gpsetdashtype{gp dt axes}
\gpsetlinewidth{0.50}
\draw[gp path] (2.640,7.573)--(6.615,7.573);
\draw[gp path] (11.027,7.573)--(11.395,7.573);
\gpcolor{color=gp lt color border}
\gpsetlinetype{gp lt border}
\gpsetdashtype{gp dt solid}
\gpsetlinewidth{1.00}
\draw[gp path] (2.640,7.573)--(2.820,7.573);
\draw[gp path] (11.395,7.573)--(11.215,7.573);
\node[gp node right] at (2.272,7.573) {$100$};
\gpcolor{color=gp lt color axes}
\gpsetlinetype{gp lt axes}
\gpsetdashtype{gp dt axes}
\gpsetlinewidth{0.50}
\draw[gp path] (2.640,1.971)--(2.640,8.133);
\gpcolor{color=gp lt color border}
\gpsetlinetype{gp lt border}
\gpsetdashtype{gp dt solid}
\gpsetlinewidth{1.00}
\draw[gp path] (2.640,1.971)--(2.640,2.151);
\draw[gp path] (2.640,8.133)--(2.640,7.953);
\node[gp node center] at (2.640,1.355) {$0$};
\gpcolor{color=gp lt color axes}
\gpsetlinetype{gp lt axes}
\gpsetdashtype{gp dt axes}
\gpsetlinewidth{0.50}
\draw[gp path] (4.391,1.971)--(4.391,8.133);
\gpcolor{color=gp lt color border}
\gpsetlinetype{gp lt border}
\gpsetdashtype{gp dt solid}
\gpsetlinewidth{1.00}
\draw[gp path] (4.391,1.971)--(4.391,2.151);
\draw[gp path] (4.391,8.133)--(4.391,7.953);
\node[gp node center] at (4.391,1.355) {$1000$};
\gpcolor{color=gp lt color axes}
\gpsetlinetype{gp lt axes}
\gpsetdashtype{gp dt axes}
\gpsetlinewidth{0.50}
\draw[gp path] (6.142,1.971)--(6.142,8.133);
\gpcolor{color=gp lt color border}
\gpsetlinetype{gp lt border}
\gpsetdashtype{gp dt solid}
\gpsetlinewidth{1.00}
\draw[gp path] (6.142,1.971)--(6.142,2.151);
\draw[gp path] (6.142,8.133)--(6.142,7.953);
\node[gp node center] at (6.142,1.355) {$2000$};
\gpcolor{color=gp lt color axes}
\gpsetlinetype{gp lt axes}
\gpsetdashtype{gp dt axes}
\gpsetlinewidth{0.50}
\draw[gp path] (7.893,1.971)--(7.893,6.352);
\draw[gp path] (7.893,7.953)--(7.893,8.133);
\gpcolor{color=gp lt color border}
\gpsetlinetype{gp lt border}
\gpsetdashtype{gp dt solid}
\gpsetlinewidth{1.00}
\draw[gp path] (7.893,1.971)--(7.893,2.151);
\draw[gp path] (7.893,8.133)--(7.893,7.953);
\node[gp node center] at (7.893,1.355) {$3000$};
\gpcolor{color=gp lt color axes}
\gpsetlinetype{gp lt axes}
\gpsetdashtype{gp dt axes}
\gpsetlinewidth{0.50}
\draw[gp path] (9.644,1.971)--(9.644,6.352);
\draw[gp path] (9.644,7.953)--(9.644,8.133);
\gpcolor{color=gp lt color border}
\gpsetlinetype{gp lt border}
\gpsetdashtype{gp dt solid}
\gpsetlinewidth{1.00}
\draw[gp path] (9.644,1.971)--(9.644,2.151);
\draw[gp path] (9.644,8.133)--(9.644,7.953);
\node[gp node center] at (9.644,1.355) {$4000$};
\gpcolor{color=gp lt color axes}
\gpsetlinetype{gp lt axes}
\gpsetdashtype{gp dt axes}
\gpsetlinewidth{0.50}
\draw[gp path] (11.395,1.971)--(11.395,8.133);
\gpcolor{color=gp lt color border}
\gpsetlinetype{gp lt border}
\gpsetdashtype{gp dt solid}
\gpsetlinewidth{1.00}
\draw[gp path] (11.395,1.971)--(11.395,2.151);
\draw[gp path] (11.395,8.133)--(11.395,7.953);
\node[gp node center] at (11.395,1.355) {$5000$};
\draw[gp path] (2.640,8.133)--(2.640,1.971)--(11.395,1.971)--(11.395,8.133)--cycle;
\node[gp node center,rotate=-270,font={\fontsize{18.0pt}{21.6pt}\selectfont}] at (0.736,5.052) {Values of $\styleparam{\tparam_1}$ and $\styleparam{\tparam_2}$};
\node[gp node center,font={\fontsize{18.0pt}{21.6pt}\selectfont}] at (7.017,0.431) {The threshold ($\styleparam{\vparam}$) of the withdrawal amount};
\draw[gp path] (6.615,6.352)--(6.615,7.953)--(11.027,7.953)--(11.027,6.352)--cycle;
\gpfill{rgb color={0.580,0.000,0.827},color=.!40} (2.640,1.971)--(2.722,1.971)--(2.722,1.971)--(2.724,1.971)%
    --(2.724,1.971)--(2.726,1.971)--(2.726,1.971)--(2.728,1.971)--(2.728,1.971)%
    --(2.729,1.971)--(2.729,1.971)--(2.731,1.971)--(2.731,1.971)--(2.733,1.971)%
    --(2.733,1.971)--(4.321,1.971)--(4.321,1.971)--(4.426,1.971)--(4.426,1.971)%
    --(4.430,1.971)--(4.430,1.971)--(11.372,1.971)--(11.372,1.971)--(11.395,1.971)%
    --(11.395,1.971)--(11.395,1.971)--(11.395,7.517)--(11.395,7.517)--(11.395,7.349)%
    --(11.372,7.349)--(11.372,7.349)--(4.430,7.349)--(4.430,4.548)--(4.426,4.548)%
    --(4.426,4.492)--(4.321,4.492)--(4.321,4.492)--(2.733,4.492)--(2.733,2.139)%
    --(2.731,2.139)--(2.731,2.083)--(2.729,2.083)--(2.729,2.027)--(2.728,2.027)%
    --(2.728,2.027)--(2.726,2.027)--(2.726,2.027)--(2.724,2.027)--(2.724,2.027)%
    --(2.722,2.027)--(2.722,2.027)--(2.640,2.027)--cycle;
\gpfill{rgb color={0.000,0.620,0.451},color=.!40} (2.640,8.133)--(2.640,2.475)--(2.722,2.475)--(2.722,2.755)%
    --(2.724,2.755)--(2.724,3.091)--(2.726,3.091)--(2.726,3.315)--(2.728,3.315)%
    --(2.728,3.820)--(2.729,3.820)--(2.729,4.828)--(2.731,4.828)--(2.731,4.940)%
    --(2.733,4.940)--(2.733,4.940)--(4.321,4.940)--(4.321,7.797)--(4.426,7.797)%
    --(4.426,7.797)--(4.430,7.797)--(4.430,7.797)--(11.372,7.797)--(11.372,8.133)--cycle;
\gpfill{color=gpbgfillcolor} (6.615,6.352)--(11.027,6.352)--(11.027,7.953)--(6.615,7.953)--cycle;
\draw[gp path] (6.615,6.352)--(6.615,7.953)--(11.027,7.953)--(11.027,6.352)--cycle;
\node[gp node right,font={\fontsize{20.0pt}{24.0pt}\selectfont}] at (8.639,7.460) {$\styleparam{\tparam_1}$};
\gpfill{rgb color={0.580,0.000,0.827},color=.!40} (9.007,7.306)--(10.659,7.306)--(10.659,7.614)--(9.007,7.614)--cycle;
\gpcolor{rgb color={0.580,0.000,0.827}}
\draw[gp path] (9.007,7.306)--(10.659,7.306)--(10.659,7.614)--(9.007,7.614)--cycle;
\gpcolor{color=gp lt color border}
\node[gp node right,font={\fontsize{20.0pt}{24.0pt}\selectfont}] at (8.639,6.844) {$\styleparam{\tparam_2}$};
\gpfill{rgb color={0.000,0.620,0.451},color=.!40} (9.007,6.690)--(10.659,6.690)--(10.659,6.998)--(9.007,6.998)--cycle;
\gpcolor{rgb color={0.000,0.620,0.451}}
\draw[gp path] (9.007,6.690)--(10.659,6.690)--(10.659,6.998)--(9.007,6.998)--cycle;
\gpcolor{color=gp lt color border}
\draw[gp path] (2.640,8.133)--(2.640,1.971)--(11.395,1.971)--(11.395,8.133)--cycle;
%% coordinates of the plot area
\gpdefrectangularnode{gp plot 1}{\pgfpoint{2.640cm}{1.971cm}}{\pgfpoint{11.395cm}{8.133cm}}
\end{tikzpicture}
%% gnuplot variables
}
% \caption{VERY GREAT PLOT}
% \label{figure:periodic_map}
\end{wrapfigure}
The PTDA in \textsc{Periodic} is shown in the right of \cref{figure:patterns}.
The PTDA matches situations where, for any two successive withdrawals of amount more than~$\vparam$, the duration between them is within $[\tparam_1, \tparam_2]$.
By the symbolic monitoring, one can identify the period of the periodic withdrawals of amount greater than $\vparam$ is in $[\tparam_1, \tparam_2]$.
An example of the validity domain is shown in the right figure.%\cref{figure:periodic_map}.

The experiment result is in \cref{figure:result:dominant-periodic}.
We observe that the execution time is linear to the number of the events and the memory usage is more or less constant with respect to the number of events.

\subsection{Discussion}

First, a positive result is that our algorithm effectively performs symbolic monitoring on more than 10,000 actions in one or two minutes even though the PTDAs feature both timing and data parameters.
The execution time in \textsc{Copy} is 50--100 times smaller than that in \textsc{Dominant} and \textsc{Periodic}.
This is because the constraint $3 < \tparam< 10$ in \textsc{Copy} is strict and the size of the configurations (\ie{} $\ObservableConf{i}$ and $\UnobservableConf{i}$ in \cref{algorithm:outline}) is small.
Another positive result is that in all of the benchmarks\LongVersion{ (\textsc{Copy}, \textsc{Dominant}, and \textsc{Periodic})}, the execution time is linear and the memory usage is more or less constant in the size of the input word.
This is because the size of configurations (\ie{} $\ObservableConf{i}$ and $\UnobservableConf{i}$ in \cref{algorithm:outline}) is bounded due to the following reason.
In \textsc{Dominant},
the loop in $\loc_1$ of the PTDA is deterministic, and 
because of the guard $\clock - \tparam_1 \in (50,100)$ in the edge from $\loc_1$ to $\loc_2$,
the number of the loop edges at $\loc_1$ in an accepting run is bounded (if the duration between two continuing actions are bounded as in the current setting).
Therefore, $|\ObservableConf{i}|$ and $|\UnobservableConf{i}|$ in \cref{algorithm:outline} are bounded.
The reason is similar in \textsc{Copy}, too.
In \textsc{Periodic}, since the PTDA is deterministic and 
the valuations of the amount of the withdrawals are in finite number,
$|\ObservableConf{i}|$ and $|\UnobservableConf{i}|$ in \cref{algorithm:outline} are bounded.

It is clear that we can design ad-hoc automata for which the execution time of symbolic monitoring can grow much faster (\eg{} exponential in the size of input word).
However, experiments showed that our algorithm monitors various interesting properties in a reasonable time.
%We note that the execution time of symbolic monitoring can grow much faster (\eg{}, exponential in the size of input word).
%However, we believe that our algorithm monitors various properties in a reasonable time.
%\ea{Suggestion for slight change: ``It is clear that we can design ad-hoc automata for which the execution time of symbolic monitoring can grow much faster (\eg{}, exponential in the size of input word).
%However, experiments showed that our algorithm monitors various interesting properties in a reasonable time.''
%}\mw{:)}

\textsc{Copy} and \textsc{Dominant} use data and timing parameters as well as memory and aggregation; from \cref{table:comparisons}, no other monitoring tool can compute the valuations satisfying the specification.
We however used the parametric timed model checker \imitator{}~\cite{AFKS12} to try to perform such a synthesis, by encoding the input log as a separate automaton; but \imitator{} ran out of memory (on a 3.75\,GiB RAM computer) for \textsc{Dominant} with $|\word| = 2000$, while \masakiTool{} terminates in 14\,s with only 6.9\,MiB for the same benchmark.
% This is probably due to the fact that a model checker keeps states in memory, while we can safely discard 
% BEGIN ICHIRO
Concerning \textsc{Periodic}, the only existing work that can possibly accommodate this specification is \cite{ADMN11}.
	While the precise performance comparison is interesting future work (their implementation is not publicly available), we do not expect our implementation be vastly outperformed: in~\cite{ADMN11}, their tool times out (after 10~min.)\ for a simple specification (``$\mathbf{E}_{[0, s_2]} \mathbf{G}_{[0, s_1]} (x < p)$'') and a signal discretized by only 128~points.

For those problem instances which \MonPoly{} and \DejaVu{} can accommodate (which are simpler and less parametrized than our benchmarks), they tend to run much faster than ours.
	For example, in \cite{HPU17}, it is reported that they can process a trace of length 1,100,004 in 30.3~seconds.
	The trade-off here is expressivity: for example, \DejaVu{} does not seem to accommodate \textsc{Dominant}, because \DejaVu{} does not allow for aggregation.
	We also note that, while \masakiTool{} can be slower than \MonPoly{} and \DejaVu{}, it is fast enough for many scenarios of real-world online monitoring.
% END ICHIRO

%%%%%%%%%%%%%%%%%%%%%%%%%%%%%%%%%%%%%%%%%%%%%%%%%%%%%%%%%%%%
%%%%%%%%%%%%%%%%%%%%%%%%%%%%%%%%%%%%%%%%%%%%%%%%%%%%%%%%%%%%
\section{Conclusion and perspectives}\label{section:conclusion}
%%%%%%%%%%%%%%%%%%%%%%%%%%%%%%%%%%%%%%%%%%%%%%%%%%%%%%%%%%%%
%%%%%%%%%%%%%%%%%%%%%%%%%%%%%%%%%%%%%%%%%%%%%%%%%%%%%%%%%%%%

\LongVersion{\paragraph{Conclusion}}
We proposed a symbolic framework for monitoring using parameters both in data and time.
Logs can use timestamps and infinite domain data, while our monitor automata can use timing and variable parameters (in addition to clocks and local variables).
In addition, our online algorithm can answer symbolically, by outputting all valuations (and possibly log segments) for which the specification is satisfied or violated.
We implemented our approach into a prototype \masakiTool{} and experiments showed that our tool can effectively monitor logs of dozens of thousands of events in a short time.

\paragraph{Perspectives}

\ea{mention that it's hard to express sequences of timed words in databases (PostGreSQL)? Or just omit it?}

Combining the BDDs used in~\cite{HPU17} with some of our data types (typically strings) could improve our approach by making it even more symbolic.
Also, taking advantage of the polarity of some parameters (typically the timing parameters, in the line of~\cite{BlT09}) could improve further the efficiency.

We \LongVersion{only }considered \emph{infinite} domains, but the case of \emph{finite} domains raises interesting questions concerning result representation:
	if the answer to a property\LongVersion{ on the log of \cref{figure:examples:copy:log}} is ``neither $\mathtt{a}$ nor $\mathtt{b}$'', knowing the domain is $\{ \mathtt{a}, \mathtt{b}, \mathtt{c} \}$, then the answer should be~$\mathtt{c}$.

\LongVersion{%
	From a usability point of view, adding some syntactic improvements to the PTDAs will help further the ease of using by non-experts\LongVersion{ (for example allowing ``$\styleact{update}(\neg \mathtt{b}, \_)$'' without guard instead of the self-loop over $\loc_0$ in \cref{figure:example:copy:automaton})}.\ea{I don't like this sentence so much :/}
}
\reviewer{3}{it may be useful to explore the use of this as a general query language for streams of data.}

% %%%%%%%%%%%%%%%%%%%%%%%%%%%%%%%%%%%%%%%%%%%%%%%%%%%%%%%%%%%%
% %%%%%%%%%%%%%%%%%%%%%%%%%%%%%%%%%%%%%%%%%%%%%%%%%%%%%%%%%%%%
% \section*{Acknowledgements}
% %%%%%%%%%%%%%%%%%%%%%%%%%%%%%%%%%%%%%%%%%%%%%%%%%%%%%%%%%%%%
% %%%%%%%%%%%%%%%%%%%%%%%%%%%%%%%%%%%%%%%%%%%%%%%%%%%%%%%%%%%%
% XXXXX

% HACK to jump to new page only for the submitted version
% \ifdefined\VersionWithComments
% \else
% 	\newpage
% \fi

\newpage

%%%%%%%%%%%%%%%%%%%%%%%%%%%%%%%%%%%%%%%%%%%%%%%%%%%%%%%%%%%%%
%%%%%%%%%%%%%%%%%%%%%%%%%%%%%%%%%%%%%%%%%%%%%%%%%%%%%%%%%%%%%

% Unlimited references: use full acronyms
	\newcommand{\CCIS}{Communications in Computer and Information Science}
	\newcommand{\ENTCS}{Electronic Notes in Theoretical Computer Science}
	\newcommand{\FMSD}{Formal Methods in System Design}
	\newcommand{\IJFCS}{International Journal of Foundations of Computer Science}
	\newcommand{\IJSSE}{International Journal of Secure Software Engineering}
	\newcommand{\JLAP}{Journal of Logic and Algebraic Programming}
	\newcommand{\JLC}{Journal of Logic and Computation}
	\newcommand{\LMCS}{Logical Methods in Computer Science}
	\newcommand{\LNCS}{Lecture Notes in Computer Science}
	\newcommand{\RESS}{Reliability Engineering \& System Safety}
	\newcommand{\STTT}{International Journal on Software Tools for Technology Transfer}
	\newcommand{\TCS}{Theoretical Computer Science}
	\newcommand{\ToPNoC}{Transactions on Petri Nets and Other Models of Concurrency}
	\newcommand{\TSE}{IEEE Transactions on Software Engineering}

\ifdefined\VersionLong
	\bibliographystyle{alpha} % alpha % plain
\else
	\bibliographystyle{splncs04} % abbrv
\fi
\bibliography{SymbMon}
%%%%%%%%%%%%%%%%%%%%%%%%%%%%%%%%%%%%%%%%%%%%%%%%%%%%%%%%%%%%%
%%%%%%%%%%%%%%%%%%%%%%%%%%%%%%%%%%%%%%%%%%%%%%%%%%%%%%%%%%%%%

\LongVersion{\newpage
\appendix
\section{Details on our algorithm for symbolic monitoring}
\label{appendix:algorithm_detail}

\paragraph{Intuition}
Intuitively, for each prefix $\word(1,i)$ of $\word$ and 
$(\tpval, \dpval) \in \PVal \times \Domain^{\VParam}$, our algorithm checks whether $\word \in \Lg(\valuate{\A}{\dtpval{\tpval}{\dpval}})$ by a breadth-first search.
However, obviously we cannot try each $(\tpval, \dpval) \in \PVal \times \Domain^{\VParam}$ because
$\PVal \times \Domain^{\VParam}$ is an infinite set.
Moreover, we have to add unobservable actions to the timed data word $\word$, where the timestamps and the number of unobservable actions are unknown.
Therefore, we symbolically represent
parameter valuations $(\tpval, \dpval) \in \PVal \times \Domain^{\VParam}$ and
concrete states $(\loc, \clockval, \varval) \in \Loc \times \CVal \times \Domain^{\Variables}$.
\ea{I don't understand this sentence :/}\mw{I updated it.}
%However, we need a symbolic representation of
%
%because 
%\begin{ienumeration}
% \item $\PVal \times \Domain^{\VParam}$ is an infinite set and 
% \item we have to add unobservable actions to $\word$ whose number and timestamps are unknown.
%\end{ienumeration}
The procedure is much like the reachability synthesis of parametric timed automata~\cite{JLR15}.
At first, we take all the parameter valuations $(\tpval, \dpval)\in \PVal \times \Domain^{\VParam}$ as candidates of $\ValidityDomain{\word}{\A}$.
Then we try each edge by a breadth-first search.
After each edge, we constrain the parameter valuations by the guards, and finally we obtain $\ValidityDomain{\word}{\A}$.

\paragraph{Notations}
In the pseudocode, we use $\clocktpvals$, $\clocktpvalsDwell$, and $\vardpvals$ for symbolic representation of valuations:
$\clocktpvals$ is a set of pairs $(\clockval,\tpval) \in \CVal\times\PVal$ of a clock valuation and a time parameter valuation;
$\clocktpvalsDwell$ is a set of triples $(\clockval,\tpval,t) \in \CVal\times\PVal\times\Rgeqzero$ of a clock valuation, a time parameter valuation, and an elapsed time; and
$\vardpvals$ is a set of pairs $(\varval,\dpval) \in \Domain^{\Variables}\times\Domain^{\VParam}$ of a variable valuation and a data parameter valuation.
We also use $\CurrConf$, $\NextConf$, and $\CurrUConf$: 
$\CurrConf$ and $\NextConf$ are finite sets of triples $(\loc,\clocktpvals,\vardpvals)$ and
$\CurrUConf$ is a finite set of triples $(\loc,\clocktpvalsDwell,\vardpvals)$, where
$\loc \in \Loc$ is a location
and
$\clocktpvals$, $\vardpvals$, and $\clocktpvalsDwell$ are as shown in the above.
% is a set of triple
%a location $\loc \in \Loc$, 
%a set $\clocktpvalsDwell \subseteq \CVal \times \PVal \times \Rnn$\LongVersion{of clock valuations, timing parameter valuations, and time elapse after the last observable action}
\ea{the last part (at least) should be in the short version; perhaps giving the role of each of these sets would be a good idea…?}\mw{I made it a little simpler.}%, and
%a set $\vardpvals \subseteq \Domain^{\Variables} \times \Domain^{\VParam}$\LongVersion{of variable valuation and data parameter valuation}.
For $\clocktpvals \subseteq \CVal \times \PVal$ and $t \in \Rnn$, 
we let $\clocktpvals + t = \{(\clockval + t,\tpval) \mid (\clockval,\tpval) \in \clocktpvals\}$.
For $\clocktpvals \subseteq \CVal \times \PVal$, $\clocktpvalsDwell \subseteq \CVal \times \PVal\times\Rgeqzero$, and $\vardpvals \subseteq \Domain^{\Variables} \times \Domain^{\VParam}$,
we denote $\project{\clocktpvals}{\TParam} = \{\tpval \mid \exists (\clockval,\tpval) \in \clocktpvals\}$,
$\project{\clocktpvalsDwell}{\TParam} = \{\tpval \mid \exists (\clockval,\tpval,t) \in \clocktpvalsDwell\}$, and
$\project{\vardpvals}{\VParam} = \{\dpval \mid \exists (\varval,\dpval) \in \vardpvals\}$.
We let $\tau_0 = 0$.

\begin{algorithm}[tbp]
 \caption{Algorithm for symbolic monitoring}
 \label{algorithm}
 %%%%% I can make this pseudo-code smaller but I did not because anyway it will be one-page
% \scalebox{1.0}{ 
% \parbox{1.0\textwidth}{
 \small
 \DontPrintSemicolon
 \newcommand{\myCommentFont}[1]{\texttt{\footnotesize{#1}}}
 \SetCommentSty{myCommentFont}
 \KwIn{A PTDA~$\A = (\ActionsWithU, \Loc, \locinit, \LocFinal, \Clock, \TParam, \Variables, \LVariables, \varvalo, \VParam, \Edges)$ over a \datatype{}~$(\Domain, \DomainGuards , \DomainUpdates)$ and actions $\ActionsWithU$,
 and a timed data word~$\word= (\action_1, \tau_1, \lvarval_1), (\action_2, \tau_2, \lvarval_2), \dots,(\action_n, \tau_n, \lvarval_n)$ over $\Actions$}
 \KwOut{$\Result$ is the validity domain $\ValidityDomain{\word}{\A}$}
 $\CurrConf \gets \{(\locinit,\{\ClocksZero\} \times \PVal ,\{\varvalo\} \times \Domain^\VParam) \};\;\Result \gets \emptyset$\; \label{alg_line:initialize}
 \For{$i \gets 1$ \KwTo $n$}{
 $\CurrUConf \gets \{(\loc,\clocktpvals \times \{0\},\vardpvals)\mid (\loc,\clocktpvals,\vardpvals) \in \CurrConf\}$\; \label{alg_line:try_unobservable_begin} \label{alg_line:set_curr_uconf}
 \tcp*{append the elapsed time from $\tau_{i-1}$}
 $\NextConf \gets \emptyset$\;
 \While (\tcp*[f]{insert $\uAction$ before $(\action_i,\tau_i,\lvarval_i)$}) {$\CurrUConf \neq \emptyset$} {
 \KwPop $(\loc,\clocktpvalsDwell,\vardpvals)$ \KwFrom $\CurrUConf$\; \label{alg_line:unobservable_pop}
 $\clocktpvalsDwell \gets \{(\clockval + d,\tpval,t + d) \mid (\clockval,\tpval,t) \in \clocktpvalsDwell, d \in \Rp.\, t + d < \tau_i - \tau_{i-1}\}$ \label{alg_line:unobservable_time_elapse}\;
 \tcp*{time elapse}
 \For (\tcp*[f]{try unobservable edges}){$\edge = (\loc, \tguard, \dguard, \uAction, \resets, \PDU, \loc') \in \Edges$}{ \label{alg_line:unobservable_for_begin}
 $\clocktpvalsDwell' \gets \{(\reset{\clockval}{\resets},\tpval,t) \mid \exists (\clockval,\tpval,t) \in \clocktpvalsDwell.\,\clockval \models \tpval(\tguard)\}$ \label{alg_line:unobservable_timed_guard} \;
 \tcp*{constrain and reset}
 $\vardpvals' \gets \{ (\DPValuate{\PDU}{\dpval}{\varval}{\lvarval_{\uAction}}, \dpval) \mid \exists (\varval,\dpval)\in \vardpvals.\, (\varval, \lvarval_{\uAction}) \models \valuate{\dguard}{\dpval} \}$ \label{alg_line:unobservable_data_guard}\;
 \tcp*{constrain and update}
 \If {$\clocktpvalsDwell' \neq \emptyset \And \vardpvals' \neq \emptyset$}{
 \KwPush $(\loc',\clocktpvalsDwell',\vardpvals')$ \KwTo $\CurrUConf$\;
 \KwPush $(\loc',\{(\clockval + \tau_i - \tau_{i-1} -  t, \tpval) \mid \exists (\clockval,\tpval,t) \in \clocktpvalsDwell'\},\vardpvals')$ \KwTo $\NextConf$\; \label{alg_line:unobservable_push_next_conf}
 \If (\tcp*[f]{found an accepting run}){$\loc'\in\LocFinal$}{
 $\Result \gets \Result \cup (\project{\clocktpvalsDwell'}{\TParam}\times\project{\vardpvals'}{\VParam})$
 \label{alg_line:unobservable_add_result}
 }
 }
 }\label{alg_line:unobservable_for_end}
 } \label{alg_line:try_unobservable_end}
 $\NextConf \gets \NextConf \cup \{(\loc,\clocktpvals + (\tau_i - \tau_{i-1}),\vardpvals) \mid (\loc,\clocktpvals,\vardpvals) \in \CurrConf\}$ \label{alg_line:try_observable_begin} \label{alg_line:observable_time_elapse}\;
 \tcp*{time elapse}
 $(\CurrConf,\NextConf) \gets (\NextConf,\emptyset)$\;
 \For (\tcp*[f]{use $(\action_i,\tau_i,\lvarval_i)$ for transition}){$(\loc,\clocktpvals,\vardpvals) \in \CurrConf$}{
 \For (\tcp*[f]{try observable edges}){$\edge = (\loc, \tguard, \dguard, \action_i, \resets, \PDU, \loc') \in \Edges$}{
 $\clocktpvals' \gets \{(\reset{\clockval}{\resets},\tpval) \mid \exists (\clockval,\tpval)\in\clocktpvals.\, \clockval \models \tpval(\tguard)\}$\;
 \tcp*{constrain and reset}
 $\vardpvals' \gets \{ (\DPValuate{\PDU}{\dpval}{\varval}{\lvarval_i}, \dpval) \mid \exists (\varval,\dpval)\in \vardpvals.\, (\varval, \lvarval_{i}) \models \valuate{\dguard}{\dpval} \}$\;
 \tcp*{constrain and update}
 \If {$\clocktpvals' \neq \emptyset \And \vardpvals' \neq \emptyset$}{
 \KwPush $(\loc',\clocktpvals',\vardpvals')$ \KwTo $\NextConf$\;
 \If (\tcp*[f]{found an accepting run}){$\loc'\in\LocFinal$}{
 $\Result \gets \Result \cup (\project{\clocktpvals'}{\TParam}\times\project{\vardpvals'}{\VParam})$
 }
 }
 }
 } \label{alg_line:try_observable_end}
 $(\CurrConf,\NextConf) \gets (\NextConf,\emptyset)$ \label{alg_line:set_next_curr_conf}
 }
 \While (\tcp*[f]{append $\uAction$ after $(\action_{n}, \tau_{n},\lvarval_{n})$}) {$\CurrConf \neq \emptyset$}{ \label{alg_line:try_final_unobservable_begin}
 \KwPop $(\loc,\clocktpvals,\vardpvals)$ \KwFrom $\CurrConf$\;
 $\clocktpvals \gets \{(\clockval + d,\tpval) \mid (\clockval,\tpval) \in \clocktpvals, d \in \Rp\}$
 \tcp*{time elapse}
 \For {$\edge = (\loc, \tguard, \dguard, \uAction, \resets, \PDU, \loc') \in \Edges$}{
 $\clocktpvals' \gets \{(\reset{\clockval}{\resets},\tpval) \mid \exists (\clockval,\tpval) \in \clocktpvals.\,\clockval \models \tpval(\tguard)\}$
 \tcp*{constrain and reset}
 $\vardpvals' \gets \{ (\DPValuate{\PDU}{\dpval}{\varval}{\lvarval_{\uAction}}, \dpval) \mid \exists (\varval,\dpval)\in \vardpvals.\, (\varval, \lvarval_{\uAction}) \models \valuate{\dguard}{\dpval} \}$\;
 \tcp*{constrain and update}
 \If {$\clocktpvals' \neq \emptyset \And \vardpvals' \neq \emptyset$}{
 \KwPush $(\loc',\clocktpvals',\vardpvals')$ \KwTo $\CurrConf$\;
 \If (\tcp*[f]{found an accepting run}){$\loc'\in\LocFinal$}{
 $\Result \gets \Result \cup (\project{\clocktpvals'}{\TParam}\times\project{\vardpvals'}{\VParam})$
 }
 }
 }
 }\label{alg_line:try_final_unobservable_end}
% }}
\end{algorithm}

\Cref{algorithm} is a pseudocode of our algorithm for symbolic monitoring.
In \cref{alg_line:initialize} of \cref{algorithm}, we set the current configurations $\CurrConf$ to be the triple $(\locinit,\{\ClocksZero\} \times \PVal ,\{\varvalo\} \times \Domain^\VParam)$, which means
we are at the initial location $\locinit$, 
the clock (resp.~variable) valuation is the initial valuation $\ClocksZero$ (resp.~$\varvalo$), and
the timing (resp.~data) parameter valuations can by any valuations $\PVal$ (resp.~$\Domain^\VParam$).
In \crefrange{alg_line:try_unobservable_begin}{alg_line:try_unobservable_end}, we try unobservable transitions.
In \cref{alg_line:set_curr_uconf}, we set the current configurations $\CurrUConf$ for the unobservable transitions, which is essentially the same as $\CurrConf$, but each $\clocktpvals$ is equipped with the time elapse after the latest observable transition.
The elapsed time $t$ is used
\begin{ienumeration}
 \item to restrict the unobservable transitions between the last observable action $\action_{i-1}$ and the next observable action $\action_{i}$ (\cref{alg_line:unobservable_time_elapse}) and
 \item to make the time elapse to $\tau_i$ (\cref{alg_line:unobservable_push_next_conf}).
\end{ienumeration}
For $(\loc,\clocktpvalsDwell,\vardpvals) \in \CurrUConf$, after time elapse in \cref{alg_line:unobservable_time_elapse}, we try unobservable edges from $\loc$ (\crefrange{alg_line:unobservable_for_begin}{alg_line:unobservable_for_end}).
We constrain the valuations $(\clocktpvalsDwell,\vardpvals)$ by the guards ($\tguard$ and $\dguard$) and conduct the reset and update in \cref{alg_line:unobservable_timed_guard,alg_line:unobservable_data_guard}.
If $(\clocktpvalsDwell,\vardpvals)$ satisfies the guards,
we add the valuations $(\clocktpvalsDwell',\vardpvals')$ and the valuations after time elapse to $\CurrUConf$ and $\NextConf$, respectively.
Moreover, if $\loc'\in\LocFinal$, we add the parameter valuations $(\project{\clocktpvalsDwell'}{\TParam},\project{\vardpvals'}{\VParam})$ to $\Result$.
After trying the unobservable edges, in \crefrange{alg_line:try_observable_begin}{alg_line:try_observable_end}, we try observable edges.
Finally, we try unobservable edges after the whole timed data word in \crefrange{alg_line:try_final_unobservable_begin}{alg_line:try_final_unobservable_end}.
The explanation of \crefrange{alg_line:try_observable_begin}{alg_line:try_observable_end} and \crefrange{alg_line:try_final_unobservable_begin}{alg_line:try_final_unobservable_end} is essentially similar to that of \crefrange{alg_line:try_unobservable_begin}{alg_line:try_unobservable_end}.

\paragraph{Termination}
Since $\A$ does not have any loop of unobservable edges, $\CurrConf$ and $\CurrUConf$ are always finite sets.
The valuations
%The clock and timing parameter valuations (and time elapse for $\clocktpvalsDwell$ and $\clocktpvalsDwell'$) 
$\clocktpvals$, $\clocktpvals'$, $\clocktpvalsDwell$, and $\clocktpvalsDwell'$ can be represented by convex polyhedra.
The time elapse (\eg{} in \cref{alg_line:unobservable_time_elapse}), 
restriction and reset (\eg{} in \cref{alg_line:unobservable_timed_guard}), and 
projection (\eg{} in \cref{alg_line:unobservable_add_result}) are standard operations on convex polyhedra and they terminate.
Therefore, if the operations on variable and data parameter valuations $\vardpvals$ and $\vardpvals'$ terminate, \cref{algorithm} terminates.

\Cref{algorithm} is correct because 
for each $\word' \in \bigl\{\word''(1,i) \mathrel{\bigl|} \project{\word''}{\Actions} = \word, i \in \{1,2,\dots,|\word''|\}\bigr\}$, 
it adds $\{(\tpval,\dpval) \mid \word' \in \Lg(\valuate{\A}{\dtpval{\tpval}{\dpval}})\}$ to $\Result$.

\begin{theorem}[correctness]
 For any PTDA~$\A$ over a \datatype{}~$(\Domain, \DomainGuards , \DomainUpdates)$ and actions $\ActionsWithU$, and for any timed data word~$\word$ over $\Actions$,
 if \cref{algorithm} terminates, we have $\Result = \ValidityDomain{\word}{\A}$ after the execution of \cref{algorithm}.
 \qed
\end{theorem}

\paragraph{Optimization}

In our implementation, we also employ an optimization to merge adjacent polyhedra in the configurations $\NextConf$ if possible. 
Precisely, we merge $(\loc,\clocktpvals,\vardpvals)$ and $(\loc',\clocktpvals',\vardpvals')$ 
in $\NextConf$ whenever we have the following:
\begin{itemize}
 \item $\loc$ and $\loc'$ are the same.
 \item $\clocktpvals$ and $\clocktpvals'$ are the same.
 \item The projection of $\vardpvals$ and $\vardpvals'$ to the valuations on strings are the same.
 \item The projection of $\vardpvals$ and $\vardpvals'$ to the valuations on integers are adjacent convex polyhedra.
\end{itemize}
Such a merge is conducted after consuming each entry $(\action_i,\tau_i,\lvarval_i)$ of the timed word $\word$ \ie{} in \cref{alg_line:set_next_curr_conf} of \cref{algorithm}.

\section{Detailed experiment results}

\Cref{table:experiment_result} shows the detailed results of our experiments.

%%%%%%%%%%%%%%%%%%%%%%%%%%%%%%%%%%
\begin{table}[h!]
 \caption{Experiment results: each cell consists of a pair $(T,M)$ of the execution time $T$ [sec.] and the memory usage $M$ [KiB] in the experiment setting.}
 \label{table:experiment_result}
 \centering
 \footnotesize
 \setlength{\tabcolsep}{2mm}
\begin{subfigure}[t]{0.33\textwidth}
 \centering
 \caption{Results of \textsc{Copy}}
 \label{table:experiment_result:copy}
 \begin{tabular}{r||c}
\multicolumn{1}{c||}{$|\word|$} & \textsc{Copy} \\ \hline\hline
4,000 & (0.66,6340) \\\hline
8,000 & (1.32,6108) \\\hline
12,000 & (2.04,6164) \\\hline
16,000 & (2.68,6168) \\\hline
20,000 & (3.45,6252) \\\hline
24,000 & (4.15,6140) \\\hline
28,000 & (4.68,6256) \\\hline
32,000 & (5.79,6256) \\\hline
36,000 & (6.14,6284) \\\hline
40,000 & (6.76,6112) \\\hline
\end{tabular}

\end{subfigure}
\begin{subfigure}[t]{0.65\textwidth}
 \centering
 \caption{Results of \textsc{Dominant} and \textsc{Periodic}}
 \label{table:experiment_result:dominant_periodic}
 \begin{tabular}{r||c|c}
\multicolumn{1}{c||}{$|\word|$} &\textsc{Dominant} & \textsc{Periodic} \\ \hline\hline
2,000 & (14.65,6928) & (6.66,6396) \\\hline
4,000 & (29.22,6964) & (14.91,6472) \\\hline
6,000 & (44.41,6964) & (16.82,6332) \\\hline
8,000 & (61.78,6956) & (27.85,6384) \\\hline
10,000 & (75.95,6936) & (36.64,6568) \\\hline
12,000 & (87.63,7032) & (37.59,6564) \\\hline
14,000 & (106.93,6984) & (55.93,6372) \\\hline
16,000 & (121.71,6948) & (57.09,6492) \\\hline
18,000 & (132.45,6952) & (61.53,6440) \\\hline
20,000 & (148.22,7236) & (69.59,6384) \\\hline
\end{tabular}

\end{subfigure} 
\end{table}
}
\end{document}